\documentclass[twocolumn]{aastex62}
\usepackage{multirow}

\usepackage{graphicx}
\newcommand\teff{$T_{\mathrm{eff}}$}

\accepted{for publication in ApJ}

\shorttitle{Towards early-type systems as extragalactic milestones. II}
\shortauthors{Taormina et al.}

\begin{document}

\title{Towards early-type eclipsing binaries as extragalactic milestones: II. NLTE spectral analysis and stellar parameters of the detached O-type system OGLE-LMC-ECL-06782 in the LMC}

\correspondingauthor{M{\'o}nica Taormina}
\email{taormina@camk.edu.pl}

\author[0000-0002-1560-8620]{Monica Taormina}
\affiliation{Centrum Astronomiczne im. Miko{\l}aja Kopernika PAN, Bartycka 18, 00-716 Warsaw, Poland}
\author{Rolf-Peter Kudritzki}
\affiliation{LMU M\"unchen, Universit\"atssternwarte, Scheinerstr. 1, 81679 M\"unchen, Germany}
\affiliation{Institute for Astronomy, University of Hawaii at Manoa, 2680 Woodlawn Drive, Honolulu, HI 96822, USA}
\author{Joachim Puls}
\affiliation{LMU M\"unchen, Universit\"atssternwarte, Scheinerstr. 1, 81679 M\"unchen, Germany}
\author{Bogumi\l Pilecki}
\affiliation{Centrum Astronomiczne im. Miko{\l}aja Kopernika PAN, Bartycka 18, 00-716 Warsaw, Poland}
\author{Eva Sextl}
\affiliation{LMU M\"unchen, Universit\"atssternwarte, Scheinerstr. 1, 81679 M\"unchen, Germany}
\author{G. Pietrzy{\'n}ski}
\affiliation{Centrum Astronomiczne im. Miko{\l}aja Kopernika PAN, Bartycka 18, 00-716 Warsaw, Poland}
\author{Miguel A.\ Urbaneja}
\affiliation{Institut f\"ur Astro- und Teilchenphysik, Universit\"at Innsbruck, Technikerstr. 25/8, 6020 Innsbruck, Austria}
\author{Wolfgang Gieren}
\affiliation{Universidad de Concepci\'{o}n, Departamento de Astronom\'{i}a, Casilla 160-C, Concepci\'{o}n, Chile}

\begin{abstract}
We combine the NLTE spectral analysis of the detached O-type eclipsing binary OGLE-LMC-ECL-06782 with the analysis of the radial velocity curve and light curve to measure an independent distance to the LMC. In our spectral analysis we study composite spectra of the system at quadrature and use the information from radial velocity and light curve about stellar gravities, radii and component flux ratio to derive effective temperature, reddening, extinction and intrinsic surface brightness. We obtain a distance modulus to the LMC of $m - M$ = 18.53 $\pm$ 0.04 mag. This value is 0.05 mag larger than the precision distance obtained recently from the analysis of a large sample of detached, long period late spectral type eclipsing binaries but agrees within the margin of the uncertainties. We also determine the surface brightnesses of the system components and find good agreement with the published surface brightness color relationship. A comparison of the observed stellar parameters with the prediction of stellar evolution based on the \texttt{MESA} stellar evolution code shows reasonable agreement, but requires a reduction of the internal angular momentum transport to match the observed rotational velocities.  
  
\end{abstract}


\section{Introduction}
One of the fundamental challenges of modern cosmology is the need for an accurate measurement of the Hubble constant better than  one percent \citep{Komatsu2011, Weinberg2013}. In this regard, double lined detached eclipsing binaries (DEBs) provide a unique way to determine distances to nearby galaxies, which can then be used as anchor points in the classical distance ladder approach. A most recent example is the determination of a 1\% precision distance (49.59 kpc) to the Large Magellanic Cloud (LMC) using 20 long period DEBs of late spectral type \citep{Pietrzynski2019}. This will provide the means to use LMC Cepheid and tip of the red giant branch stars towards a 1 \% determination of H$_{0}$ and may lead to stronger evidence for physics beyond the standard cosmological $\Lambda$CDM model \citep{Riess2019, Freedman2019}. The fundamental advantage of using late-type DEBs is that their surface brightness is very precisely determined by their V-K color through the surface brightness - color relationship, which can be measured empirically through long-baseline interferometry and has a scatter smaller than 1\%. Thus, the angular diameters of such systems can be obtained simply from a measurement of their color. The combination with the stellar radii derived from an analysis of radial velocity and eclipse light curves then yields a straightforward geometrical distance, typically with an accuracy of the order of 2\%.

The downside of using late-type DEBs for extragalactic distance determinations is that these systems are faint, typically m$_V \sim$ 18.5 at the distance of the LMC, which makes it impossible to use them for galaxies significantly more distant. On the other hand, DEBs of early spectral type are much brighter and can be easily used out to distances of 1 Mpc, for instance for galaxies such as M33 and M31 \citep{Ribas2005, Bonanos2006, Vilardell2010}. However, the problem of distances obtained in this way is that they rely on surface brightnesses predicted by model atmosphere theory, once the effective temperatures of the components have been determined by quantitative spectroscopy or spectrophotometry. The atmospheres of hot massive stars are affected by strong deviations from local thermodynamic equilibrium, severe metal line blanketing and by the hydrodynamics of stellar winds. All these effects need to be carefully taken into account and, while there has been enormous progress over the last decades in modelling the atmospheres of hot massive stars, the jury is still out how accurate the surface brightness predictions of the different available model atmosphere codes are. In addition, massive early type stars, just born in regions of heavy star formation, are usually severly affected by interstellar reddening with reddening laws that deviate from the standard reddening law usually applied \citep{Urbaneja2017, Maiz2014, Maiz2017}. This adds to the complexity of distance determination using early-type DEBs.  In the case of the LMC distance moduli obtained in this way range from 18.20 to 18.55 mag \citep{Udalski1998, Guinan1998, Ribas2000, Nelson2000, Fitzpatrick2002, Ribas2002, Fitzpatrick2003, Bonanos2011} indicating an uncertainty much larger than the 1\% distance obtained from late-type DEBs.

In view of this situation we have started a comprehensive project to re-investigate the use of early-type DEBs for accurate distance determinations. The goal is to study a large number of systems in the LMC, where the distance is now extremely well known, to empirically derive surface brightnesses for hot DEBs and to improve the calibration of the surface brightness- color relationship on its extension towards hot stars. In addition, the accuracy of distances determined with early type DEBs can be tested directly. First results have been presented recently by \citeauthor{Taormina2019} (\citeyear{Taormina2019}; Paper I) based on a light curve and radial velocity analysis of two systems in the LMC, OGLE-LMC-ECL-22270 (BLMC-01) and OGLE-LMC-ECL-06782 (BLMC-02). BLMC-01 is a relatively cool system consisting of two evolved B-stars, where the analysis allowed to constrain masses, radii, temperature and reddening in a straightforward way by utilizing information from spectral type and photometry in addition to light and radial velocity curves. In particular a slight oblateness of one of the components has weakened the degeneracy and let us determine the radii very precisely.

The case of BLMC-02, on the other hand, turned out to be more complicated. Both components in this system are O-stars and, thus, the determination of reddening and extinction just based on photometry and spectral type is highly uncertain. In addition, inclination angle and stellar radii are not as well constrained from the light curve analysis with stellar radii uncertain by 2 to 3 percent, due to the lack of independent spectroscopic flux ratios needed in case of not totally eclipsing, approximately spherical stars.
An improvement regarding the above mentioned problems can only come from a detailed quantitative spectroscopic analysis based on non-LTE model atmospheres. In this paper, we carry out such an analysis and present new results with respect to stellar temperatures and radii, reddening and extinction, surface brightnesses and intrinsic colors and we compare the distance obtained with the model atmosphere fitting method with the well constrained LMC distance. We also discuss the status of the system with respect to stellar evolution.

 \begin{figure*}[ht!]
  \begin{center}
\includegraphics[scale=0.40]{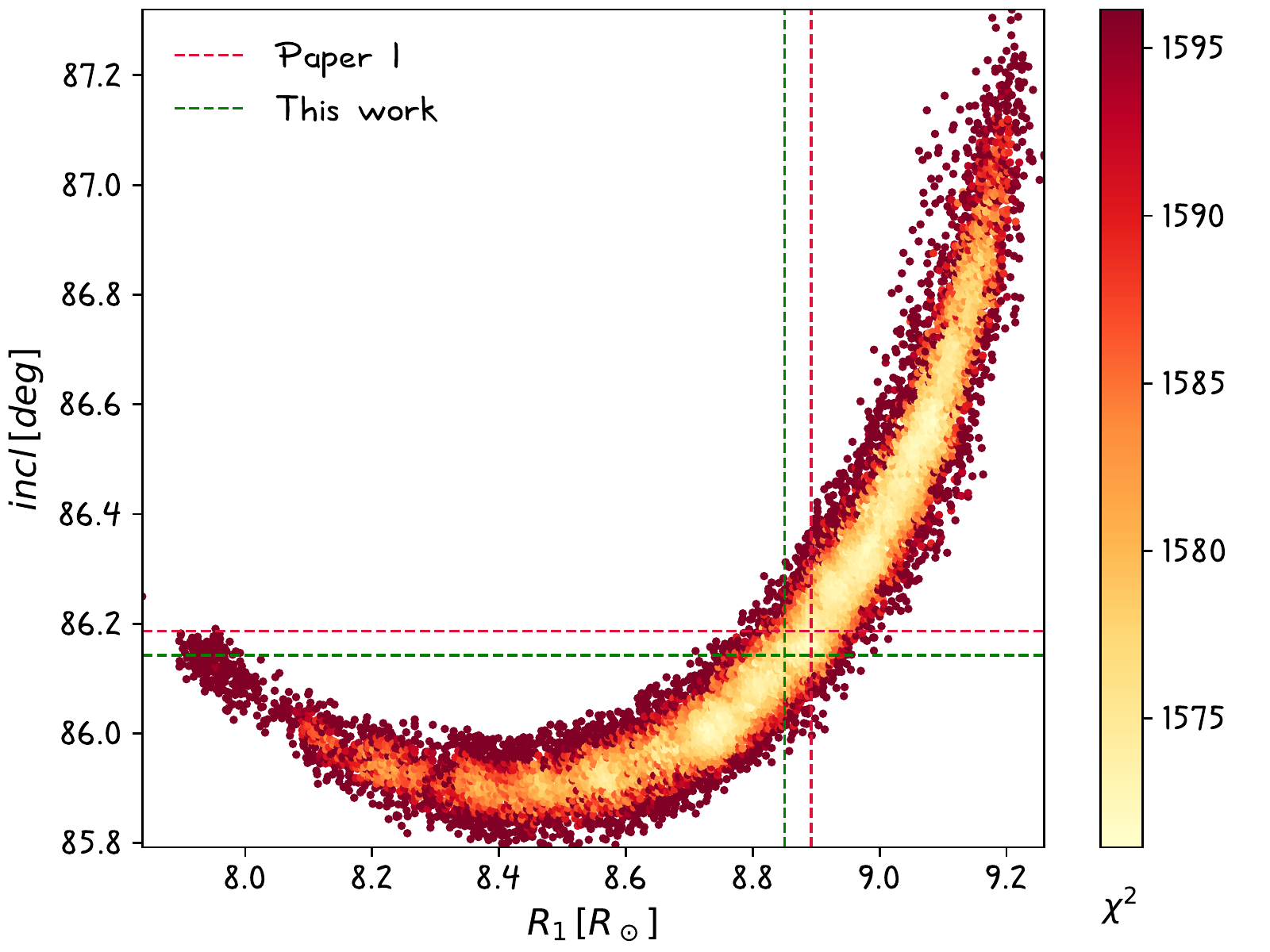}
\includegraphics[scale=0.40]{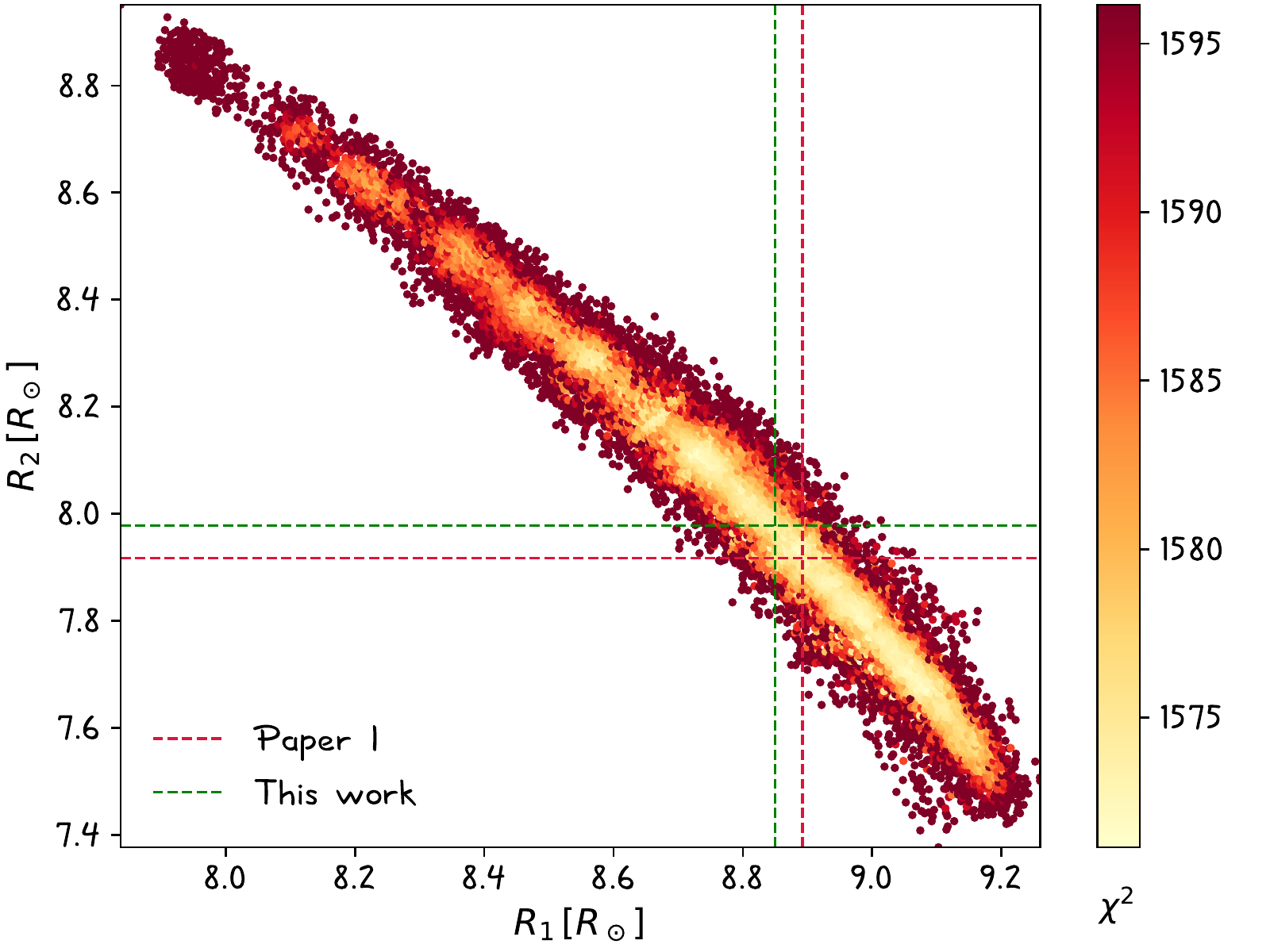}  
\includegraphics[scale=0.40]{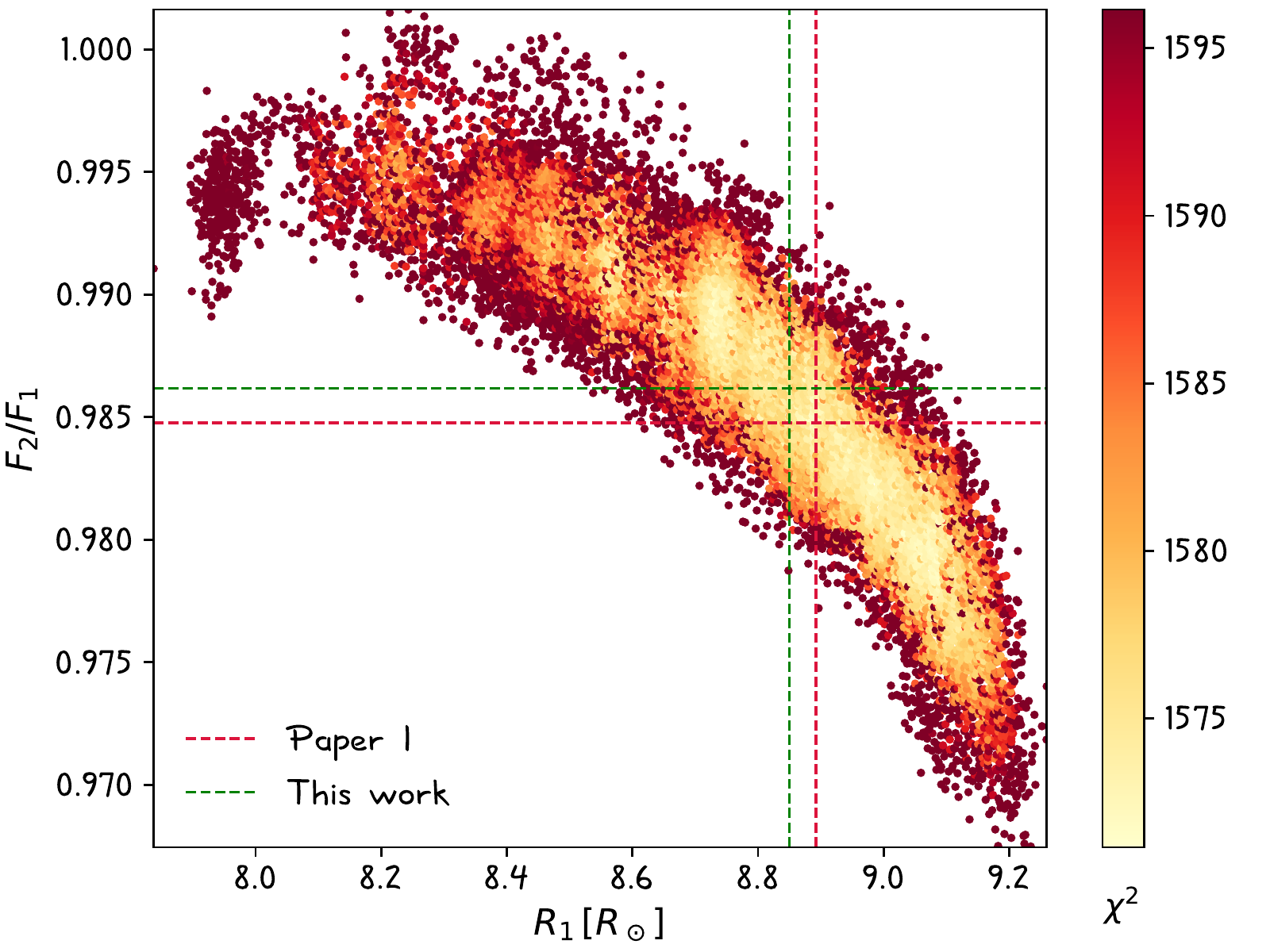}
  \end{center}
  \caption{
	  Monte Carlo light curve analysis of BLMC-02 displaying individual solutions for inclination angle (upper left), secondary stellar radius R$_2$ (upper right), and the surface brightness ratio of the two components in the V-band (bottom) versus primary stellar radius R$_1$. The solutions are color coded by their $\chi^2$ values. The color code is given by the side bars. The vertical and horizontal green dashed lines mark the final values obtained by the spectroscopic analysis. For a comparison, we include the solution adopted in Paper I (red dashed lines). } \label{fig:radii}
\end{figure*}

\begin{figure}[t]
\includegraphics[scale=0.40]{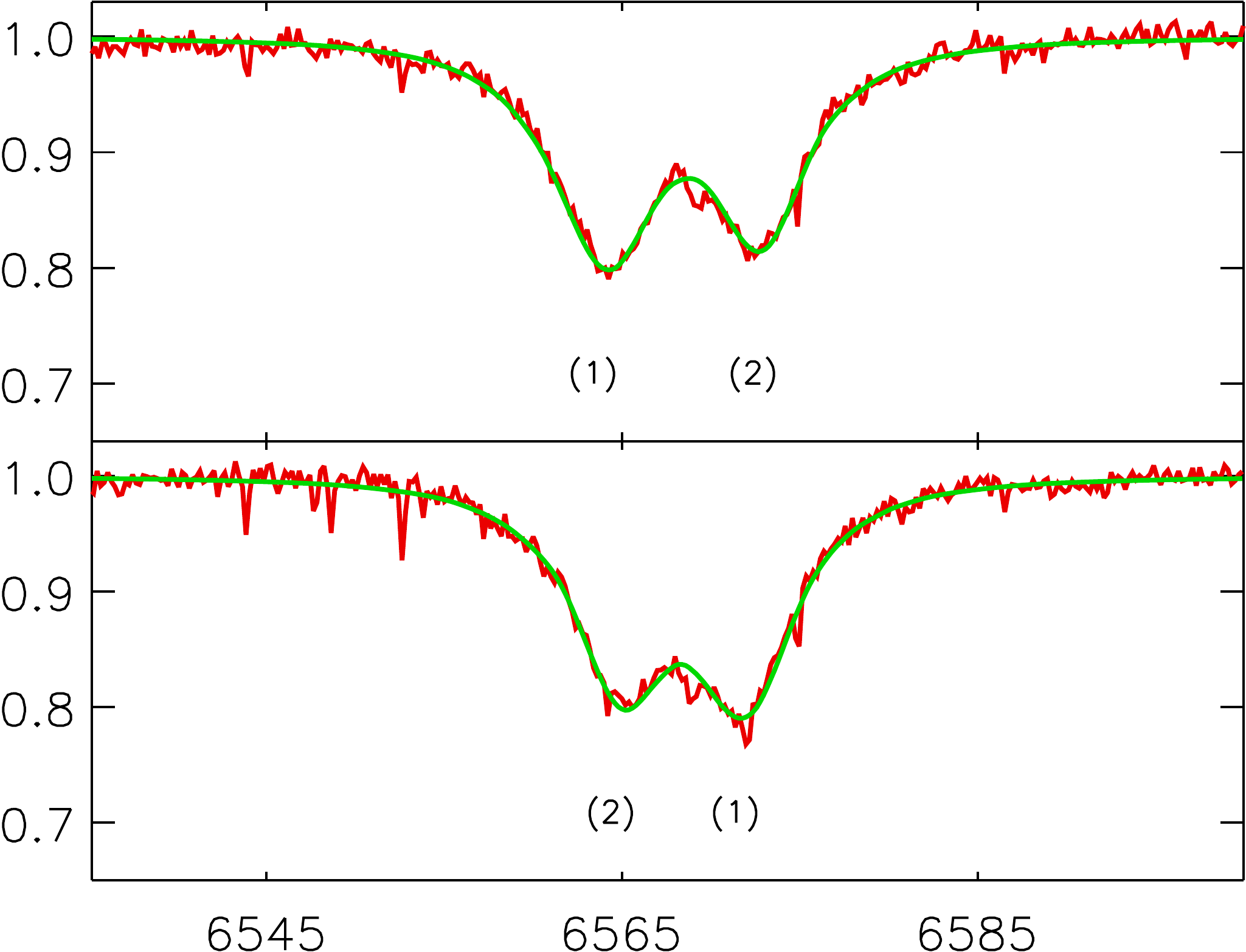}
\caption{Fit of the hydrogen line H$_{\alpha}$ of the UVES1 (top) and UVES2 (bottom) spectrum. The line contributions of the primary and secondary are indicated by (1) and (2), respectively.}
\label{fig:halpha}
\end{figure}

\section{Observations}
	The spectra available for our analysis were obtained with the high resolution spectrographs UVES \citep{Dekker2000} and MIKE \citep{Bernstein2003} attached to the ESO VLT and the Magellan Clay Telescope, respectively.  UVES observations were taken with the standard configuration DIC1 390+564, which provides a coverage of two wavelength ranges: 3260–4520 \AA \, and 4580–6690 \AA, while the MIKE configuration provides two overlapping spectra, that cover a wavelength range from 3200 to 10000 \AA. More details on the collected data can be found in Paper I.

Our analysis method uses the composite spectra of both components at binary phases around quadrature, where the radial velocity shifts are large enough so that the lines of the two components are well separated. To increase the S/N ratio all spectra were binned and then sampled to a resolution of 0.2 \AA. In addition, to improve the S/N for the analysis we added three UVES spectra at phases 0.13, 0.16, and 0.18 (spectrum ``UVES1'') and  0.64, 0.87, and 0.90 (spectrum ``UVES2'') as well as four MIKE spectra at phases 0.17, 0.21, 0.32, and 0.35 (spectrum ``MIKE'') to obtain three spectra with very good S/N of the order of 150. The relative radial velocity changes between the individual phases for the added spectra are small compared to the spectral line widths which are dominated by the high rotational velocities of 105 km/s. The radial velocity spread of both sets of UVES spectra amounts to $\sim \pm$ 15 km/s and to $\sim \pm$ 25 km/s for the MIKE spectra. These three sets of co-added and normalized spectra - UVES1, UVES2, and MIKE - form the basis for our spectroscopic analysis.

\begin{figure*}[t]
  \includegraphics[width=\textwidth]{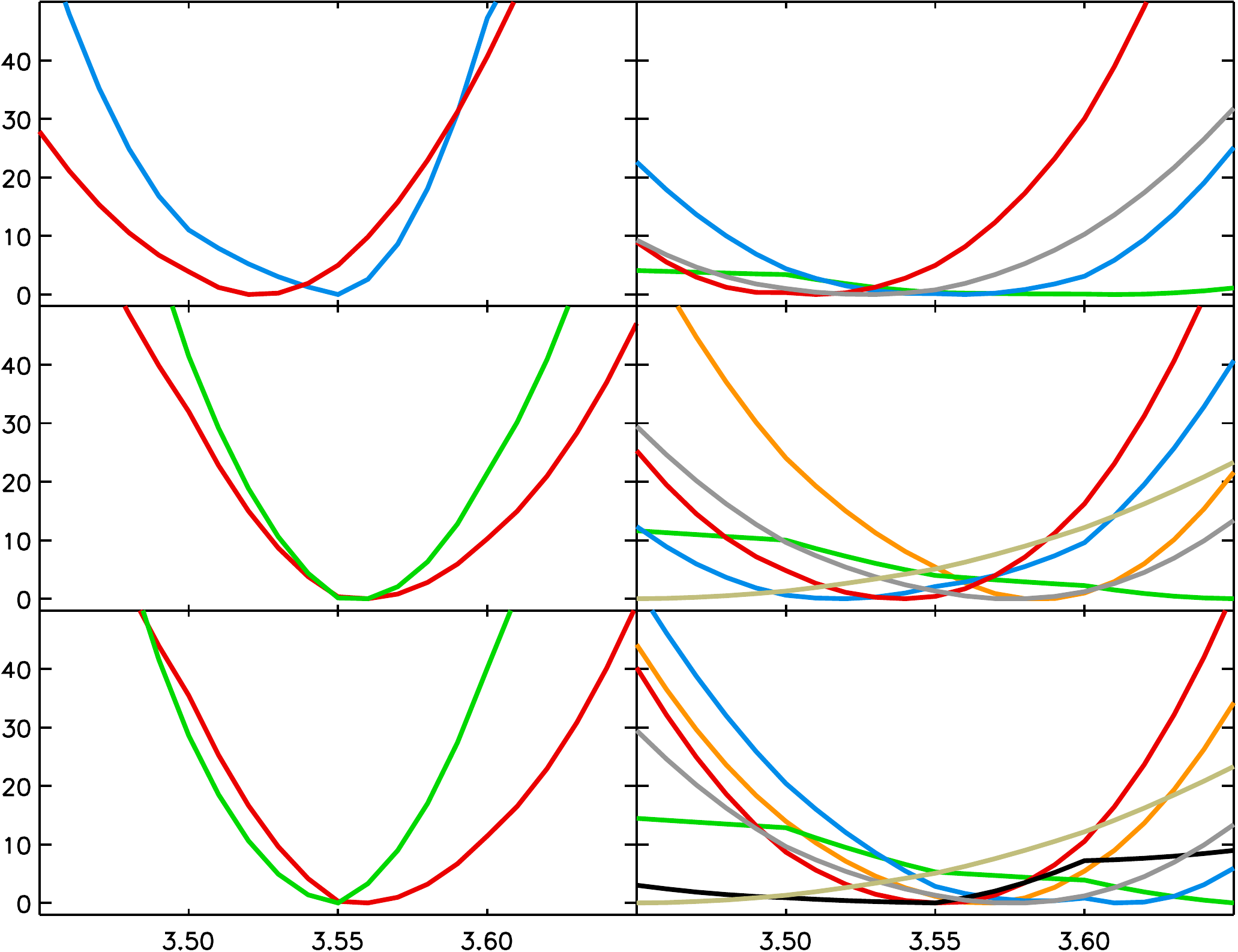}
\caption{$\Delta \chi^2$ = $\chi^2$(T$_ 1$) - $\chi^2_{min}$ as a function of primary temperature T$_1$/10$^4$K for the diagnostic helium lines and at a fixed set of radii, suface brightness ratio and gravities corresponding to R$_1$ = 8.85 R$_{\odot}$. Left: ionized helium lines (red: HeII4200, blue: HeII 4542, green: HeII 5411); right: neutral helium lines (black: HeI 4026, grey: HeI 4143, red: HeI 4387, blue: HeI 4471, green: HeI 4713, orange: HeI 4922, gold: HeI 5048. Upper row: MIKE spectrum, middle row: UVES1, bottom row: UVES2.}
\label{fig:chi_lines}
\end{figure*}

    For determination of interstellar reddening and extinction we need multi-band photometry of the system outside the eclipse. From Paper I we have the Johnson-Cousins apparent magnitudes V = 13.715 $\pm$ 0.02 mag, R = 13.764 $\pm$ 0.035 mag, and I$_C$ = 13.912 $\pm$ 0.02 mag. 

    The system has also been observed in the \emph{Gaia} mission \citep{Gaia2016}. For our purpose, we used the mean apparent brightness of the system in the G$_{BP}$ (330-680 $nm$) and G$_{RP}$ (630-1050 $nm$) bands published in the second \emph{Gaia} data release \citep[DR2,][]{Gaia2018}. The catalogue is based on 22 months of observations between July 2014 and May 2016. During this period, 31 observations were carried out and from the dates, when the field was observed, we can work out a correction for the published mean magnitudes to recover the photometry outside the eclipses. The correction is -0.055 mag and we obtain the blue and red magnitudes G$_B$ = 13.481 $\pm$ 0.020 mag and G$_R$ = 13.831 $\pm$ 0.020 mag. 

    In addition, near-IR 2MASS 6X photometry \citep{Cutri2012} is available for BLMC-02. As can be inferred from the timing, the 2MASS 6X observations were carried out at the edge of the secondary eclipse at phase 0.4396, resulting in a small correction of $\Delta$m = -0.009 mag calculated using our light curve model of Paper I. With this correction  we obtain the near-IR magnitudes outside eclipse, J = {14.169 $\pm$ 0.023 mag, H = 14.267 $\pm$ 0.028 mag, and K$_S$ = 14.335 $\pm$ 0.036 mag.

    For the purpose of the subsequent spectroscopic analysis we checked the effect of the mutual irradiation (known also as the reflection effect) on the stars. Although the separation is relatively small, the temperatures are very similar, so we did not expect a strong effect. By turning on and off the reflection in the modeling code we found that the difference between the models close to the eclipses (i.e. where it should be the strongest) was only about 0.1 mmag.  We consider this effect as neglibible and we do not take it into account in the analysis.

\section{Stellar Radii and Surface Brightness Ratios}

\begin{figure}[t]
\includegraphics[scale=0.45]{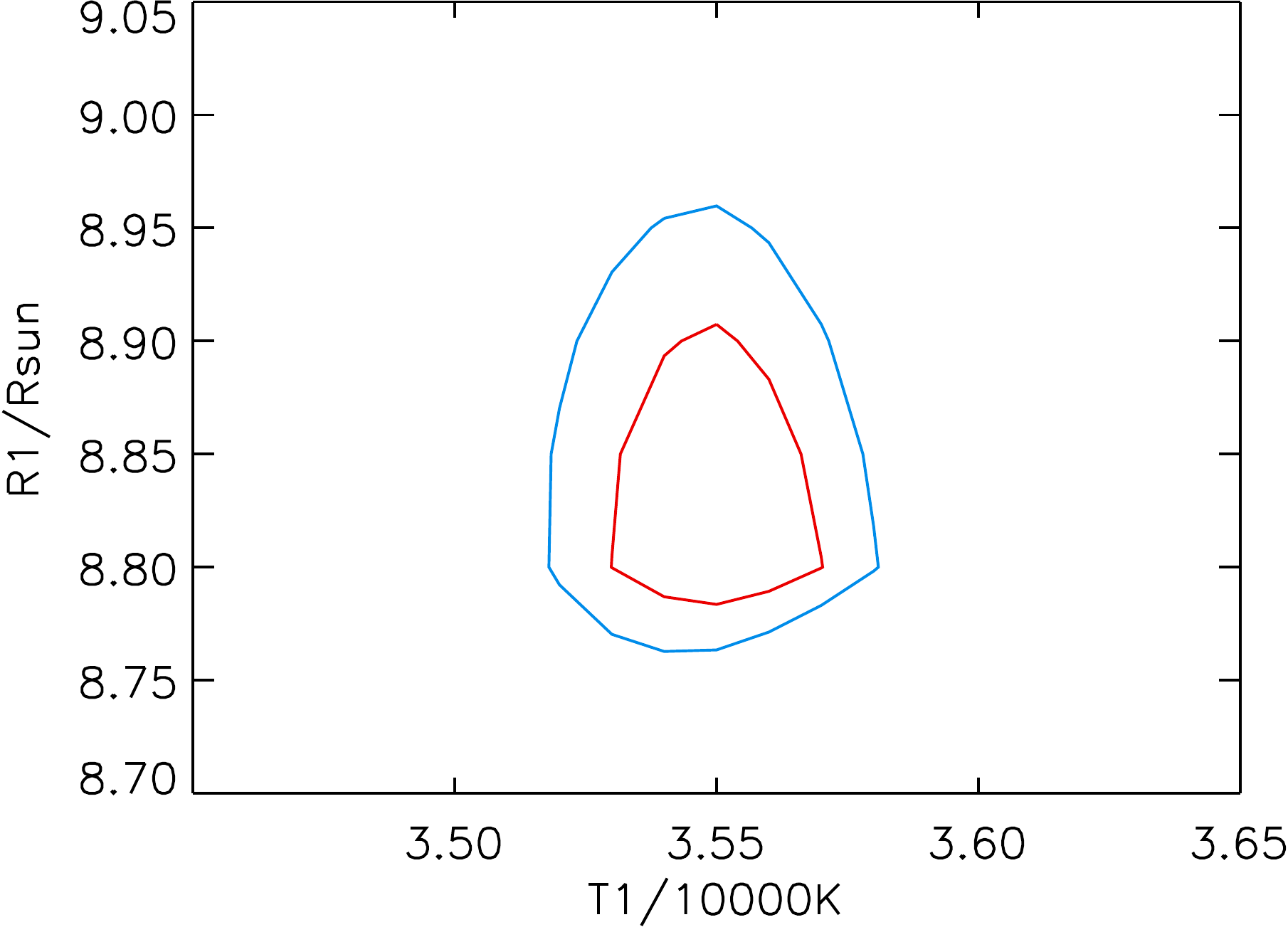}
\caption{Isocontours $\Delta \chi^2$ = $\chi^2$(R$_1$, T$_ 1$) - $\chi^2_{min}$ in the (R$_1$, T$_1$)-plane. When compared with extensive Monte-Carlo simulations of the line fitting process the red isocontour encompasses 68\% of the solutions and the blue isocontour covers 95\%.}
\label{fig:chisq}
\end{figure}

 \begin{figure*}[ht!]
  \begin{center}
\includegraphics[scale=0.50]{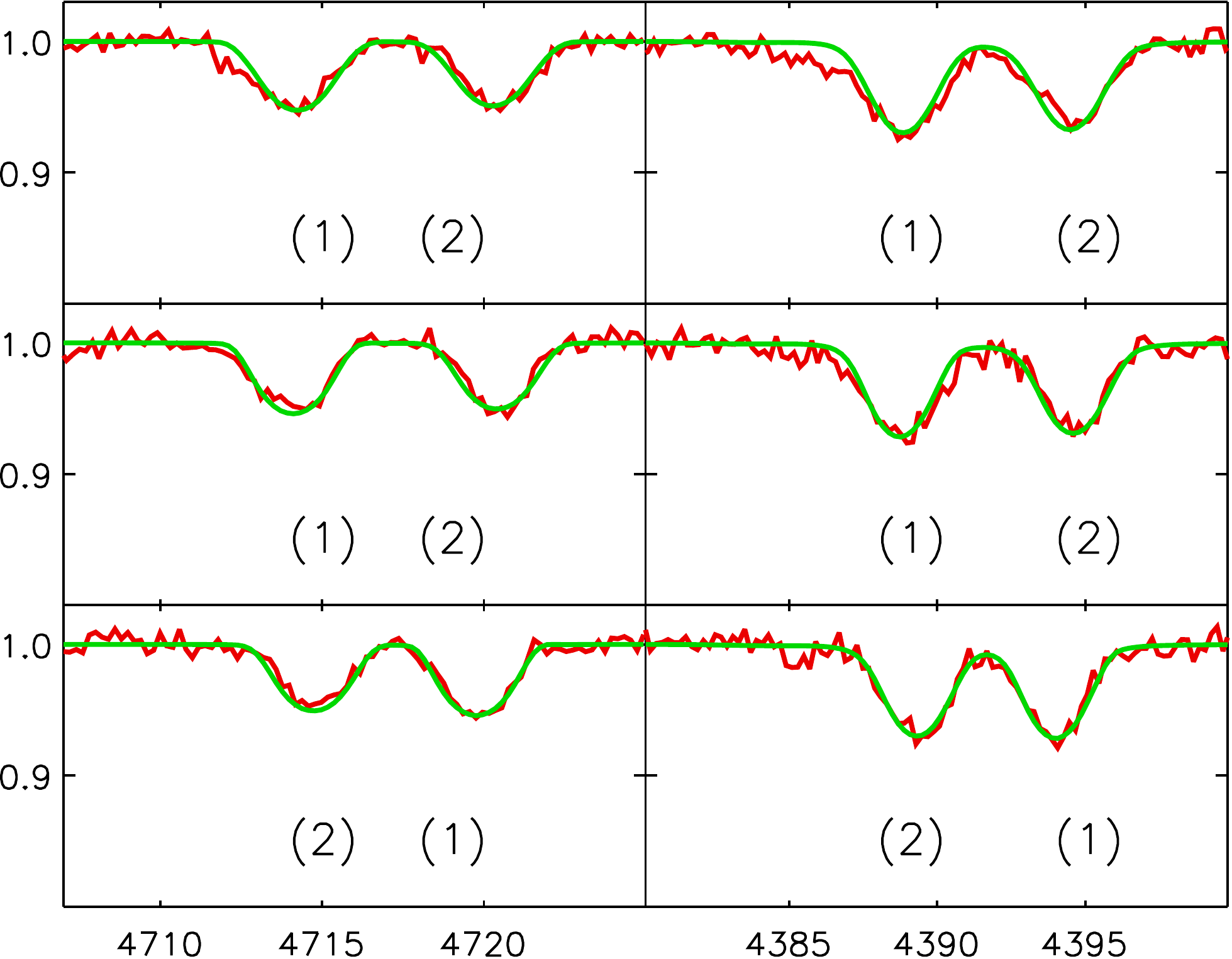}
  \end{center}
  \caption{
Fit of the HeI 4713 (left) and HeI 4387 (right) of the MIKE (top), UVES1 (middle) and UVES2 (bottom) spectra with the best composite binary model spectrum obtained from the $\chi^2$-analysis. The line contributions from the primary and secondary in each spectrum are indicated by (1) and (2), respectively.} \label{fig:He_1}
\end{figure*}

	The spectroscopic analysis of the composite spectrum requires knowledge of the stellar radii of the two components and their surface brightness ratios F$_2$/F$_1$, which are usually well constrained through the light curve analysis. However, in the case of BLMC-02 we encounter a slight degeneracy between inclination angle, stellar radii and surface brightness ratio. This is demonstrated in Figure \ref{fig:radii}, which displays the results of a Monte Carlo analysis of the light curves. Although the sum of the radii is determined with a precision of 0.5\% (i.e. $R_1+R_2 = 16.81 \pm 0.08 \, R_\odot$) due to the degeneration the individual radii are precise only to 2-3\%.
The light curve analysis allows a range of primary radius R$_1$ between 8.75 and 9.15 R$_{\odot}$ with secondary radii R$_2$ between 8.10 and 7.56 R$_{\odot}$, while the surface brightness ratio ranges between 0.989 and 0.976. Note that the analysis of the light curve in paper I reveals a small oblateness $\Delta$ = (R$_{max}$ -R$_{min}$)/R$_{max}$ = 0.04 for both components. As in Paper I, we use the equivalent radius, which is the radius of a sphere with the same volume as the oblate star, to characterize stellar radii.

\begin{figure}[ht!]
  \begin{center}
    \includegraphics[scale=0.35]{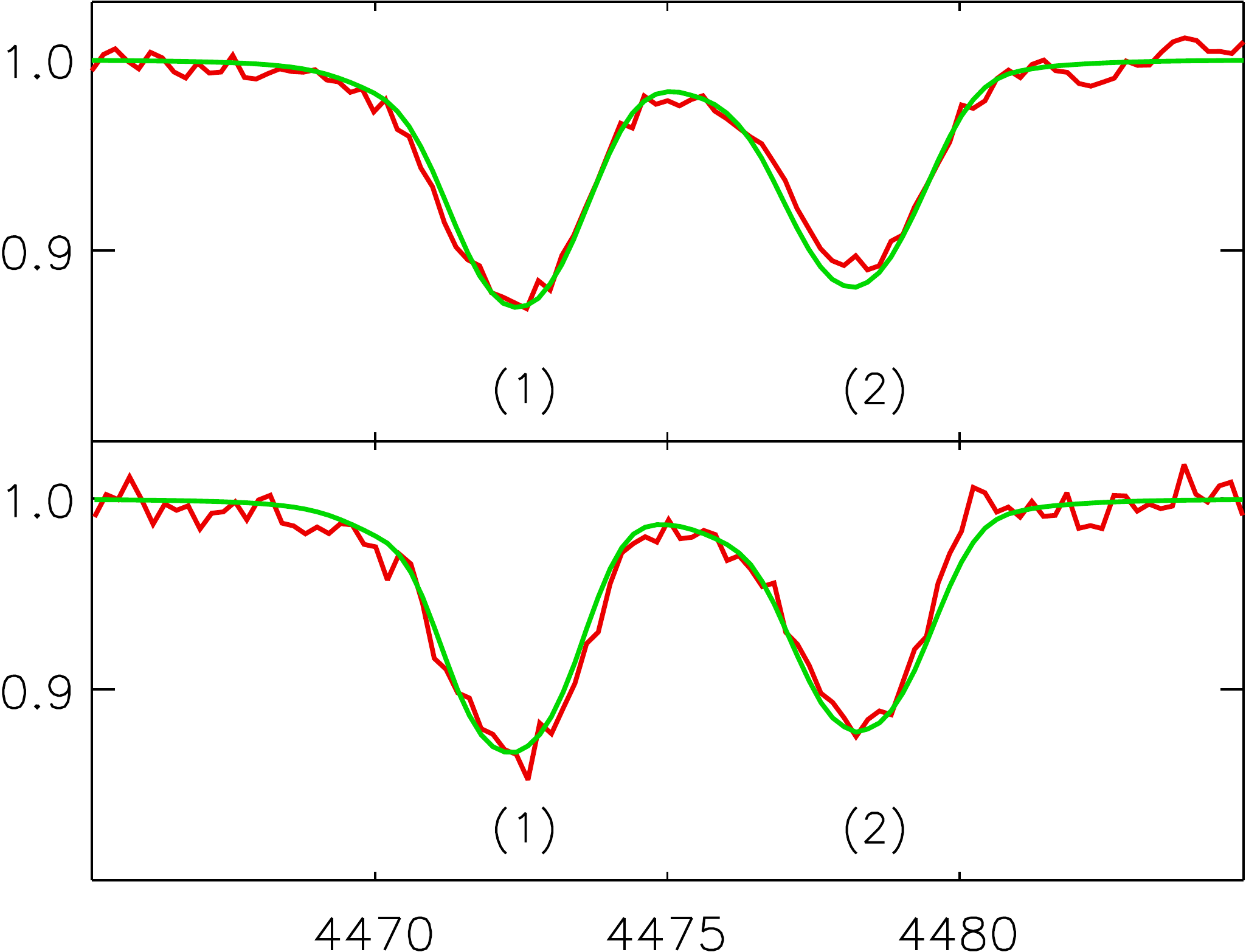}
    \includegraphics[scale=0.37]{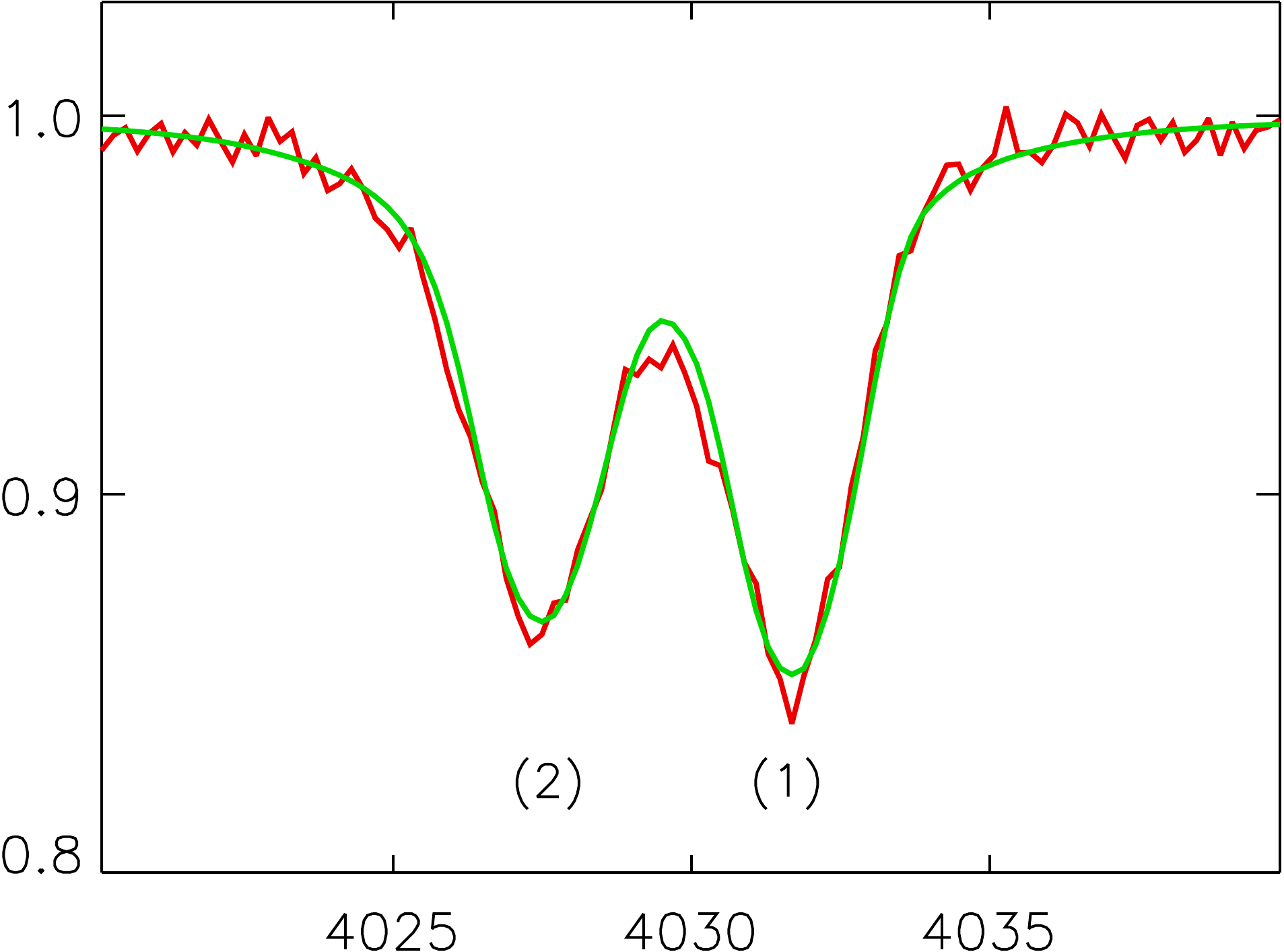}
    \includegraphics[scale=0.40]{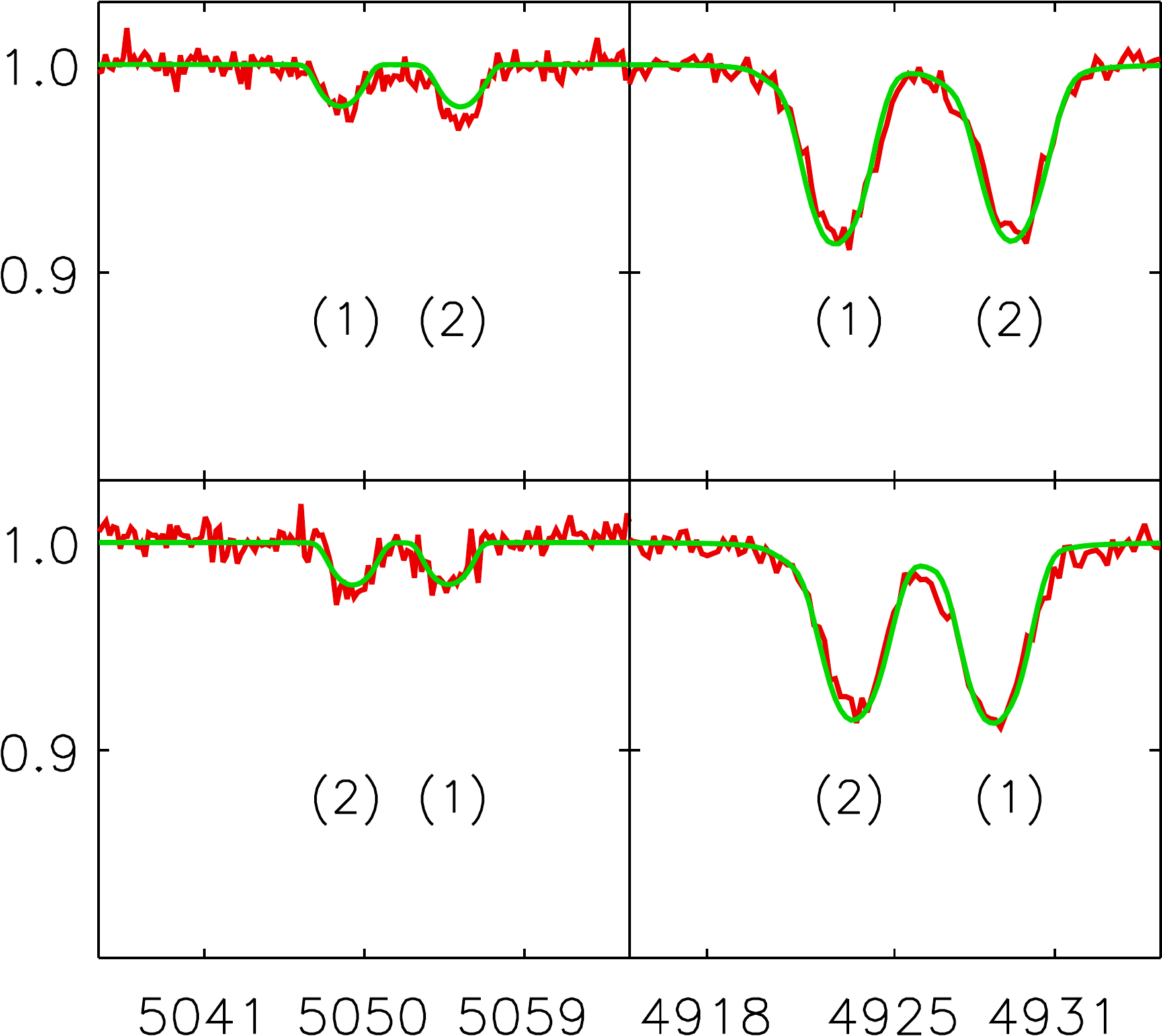} 
  \end{center}
  \caption{
Fit of HeI 4471 (upper figure:  MIKE(top), UVES1(bottom)), HeI 4026 (middle figure, UVES2), HeI 5048 (bottom left, Top: UVES1, bottom: UVES2), and HeI 4922 (bottom right, Top: UVES1, bottom: UVES2)} \label{fig:He_2}
\end{figure}

\section{Spectroscopic Analysis}

\subsection{Model atmosphere calculations}

For our model atmosphere analysis we use the non-LTE model atmosphere code \texttt{FASTWIND} \citep{Puls2005, Rivero2012a} to calculate normalized composite spectra and energy distributions (SEDs).  \texttt{FASTWIND}  has been developed to model the spectra of hot stars. It includes the important effects of stellar winds, atmospheric spherical extension and non-LTE metal line blanketing. We calculate normalized flux spectra F$_{1,\lambda}$ and F$_{2,\lambda}$ for the primary and secondary star, respectively, and then combine those through (see \citealt{Bonanos2006})

\begin{equation}
  F_{\lambda} = w_1F_{1,\lambda}(T_1, log~g_1, v_1) + w_2F_{2,\lambda}(T_2, log~g_2, v_2)
\end{equation}

where $T_i$, $log \, g_i$ and $v_i$ denote effective temperature, gravity and radial velocity of the primary and secondary. In the calculation of the spectra the broadening through stellar rotation is taken into account. The weights $w_i$ are calculated as

\begin{equation}
w_1 = {1 \over 1 + (R_2/R_1)^2F_2/F_1}, w_2 = 1 - w_1 .
\end{equation} 

To produce model spectra, which can be compared with the three co-added observed UVES1, UVES2, MIKE spectra we calculated spectra with Doppler shifts corresponding to the phases of the primary and secondary radial velocity curves, where the individual spectra were taken, and then co-added the individual model spectra to exactly simulate the way how the observed spectra were obtained. 

With the results from the previous subsection we define the grid of radii, surface brightnesses and gravities given in Table \ref{tab:radtab}. Note that the gravities follow from the masses of the two components, which are very precisely constrained by the analysis of the radial velocity curve in Paper I.  We then use the values in Table \ref{tab:radtab} to calculate composite spectra as a function of the primary stellar radius. Of course, this also requires a choice of effective temperatures. We can use the observed surface brightness ratios of Table \ref{tab:radtab} to constrain the effective temperature difference $\Delta$T = T$_1$ - T$_2$ between the two components. With the  \texttt{FASTWIND}  model SEDs we can calculate stellar fluxes in the V-band for the primary and secondary and then their ratio. In the \teff~range between 32000K and 38000K expected from the spectral types of the two components (both objects have the spectral type O7.5) the observed surface brightness ratios correspond to $\Delta$T = $1000\pm100$ K. This leaves the primary temperature T$_1$ as the only other free parameter besides the choice of the primary radius R$_1$.

   \begin{figure*}[ht!]
  \begin{center}
\includegraphics[scale=0.50]{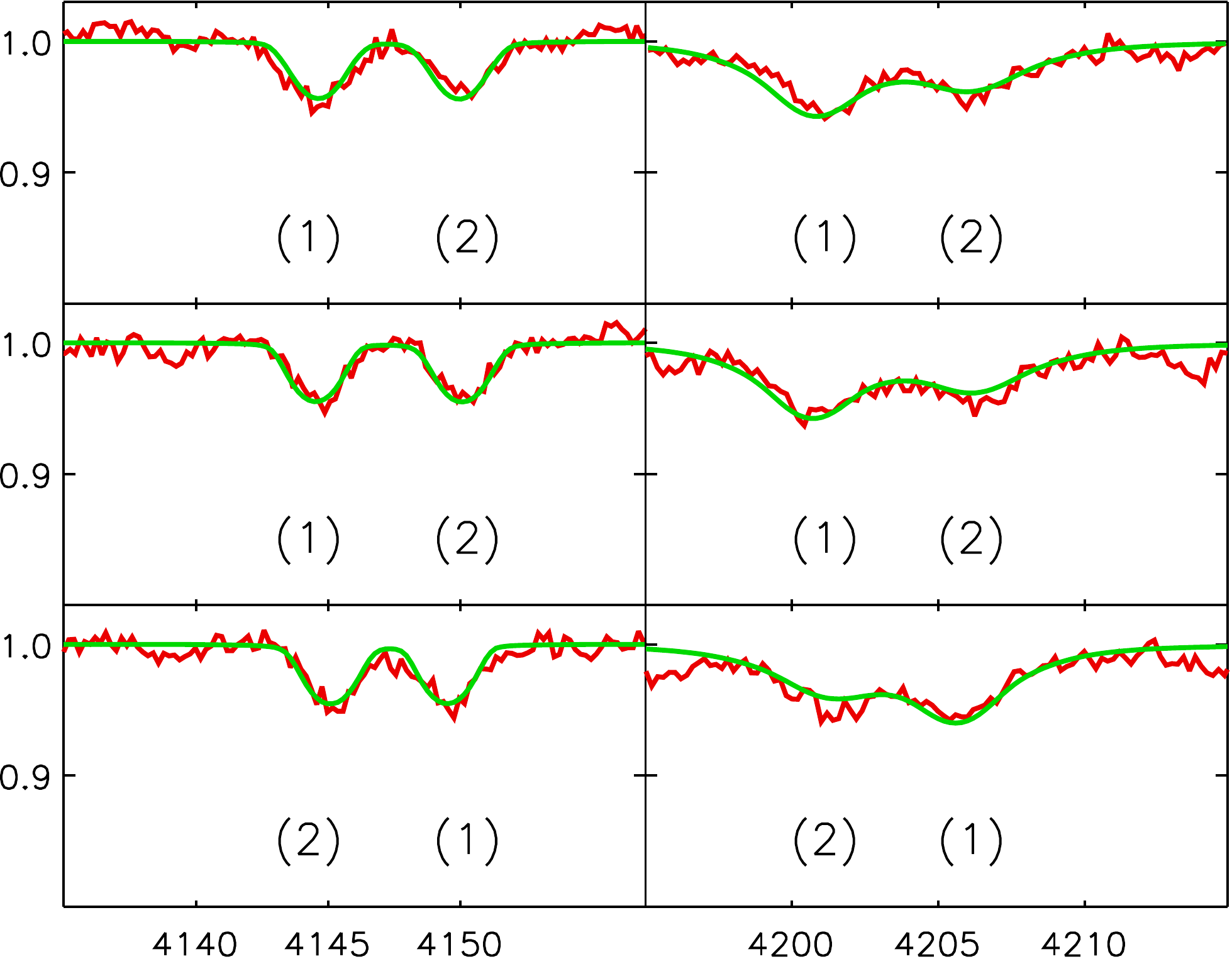}
  \end{center}
  \caption{
Fit of HeI 4143 (left) and HeII 4200 (right). MIKE spectrum at the top, UVES1 spectrum in the middle and UVES2 at the bottom.} \label{fig:He_3}
\end{figure*}

We also need to make a choice for metallicity and helium abundance. Following the work by \cite{Urbaneja2017}, in which metallicities of a large sample of young massive hot stars in the LMC were studied spectroscopically, we adopt a metallicity of [Z]=log Z/Z$_{\odot}$ = -0.35. Since we will use the helium ionization equilibrium, i.e. the observed strengths of line profiles of neutral and ionized helium lines, to determine the two free parameters R$_1$ and T$_1$, we also need to adopt a helium abundance. The standard technique to constrain the helium abundance is to adopt the value where HeI and HeII lines fit at the same effective temperature. For this purpose, we have adopted four values of the helium abundance N(He)/N(H), 0.08, 0.09, 0.1, 0.15, for which we carried out the fit of the helium lines as described below.

\begin{figure*}[ht!]
  \begin{center}
    \includegraphics[scale=0.35]{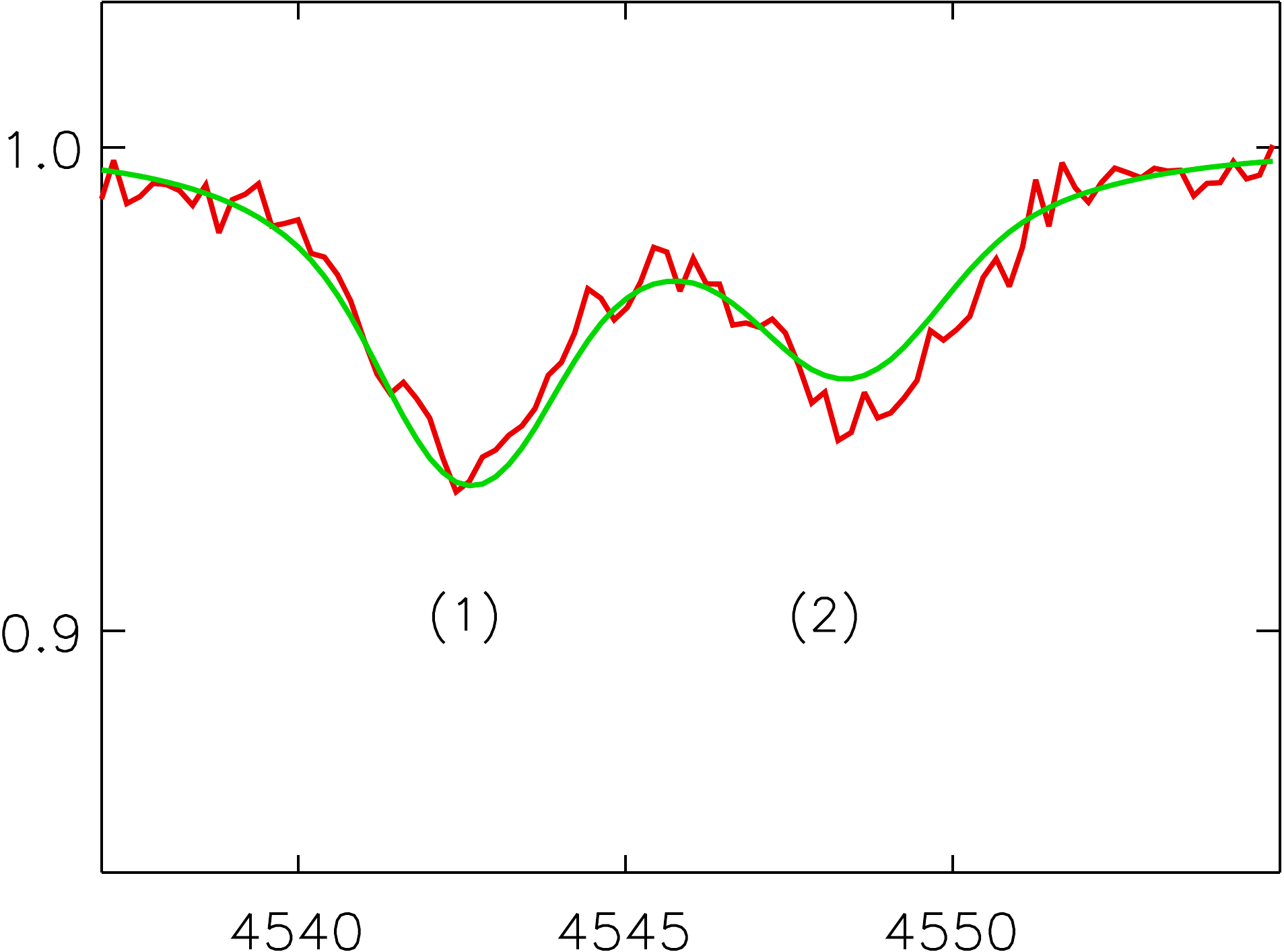}
    \includegraphics[scale=0.35]{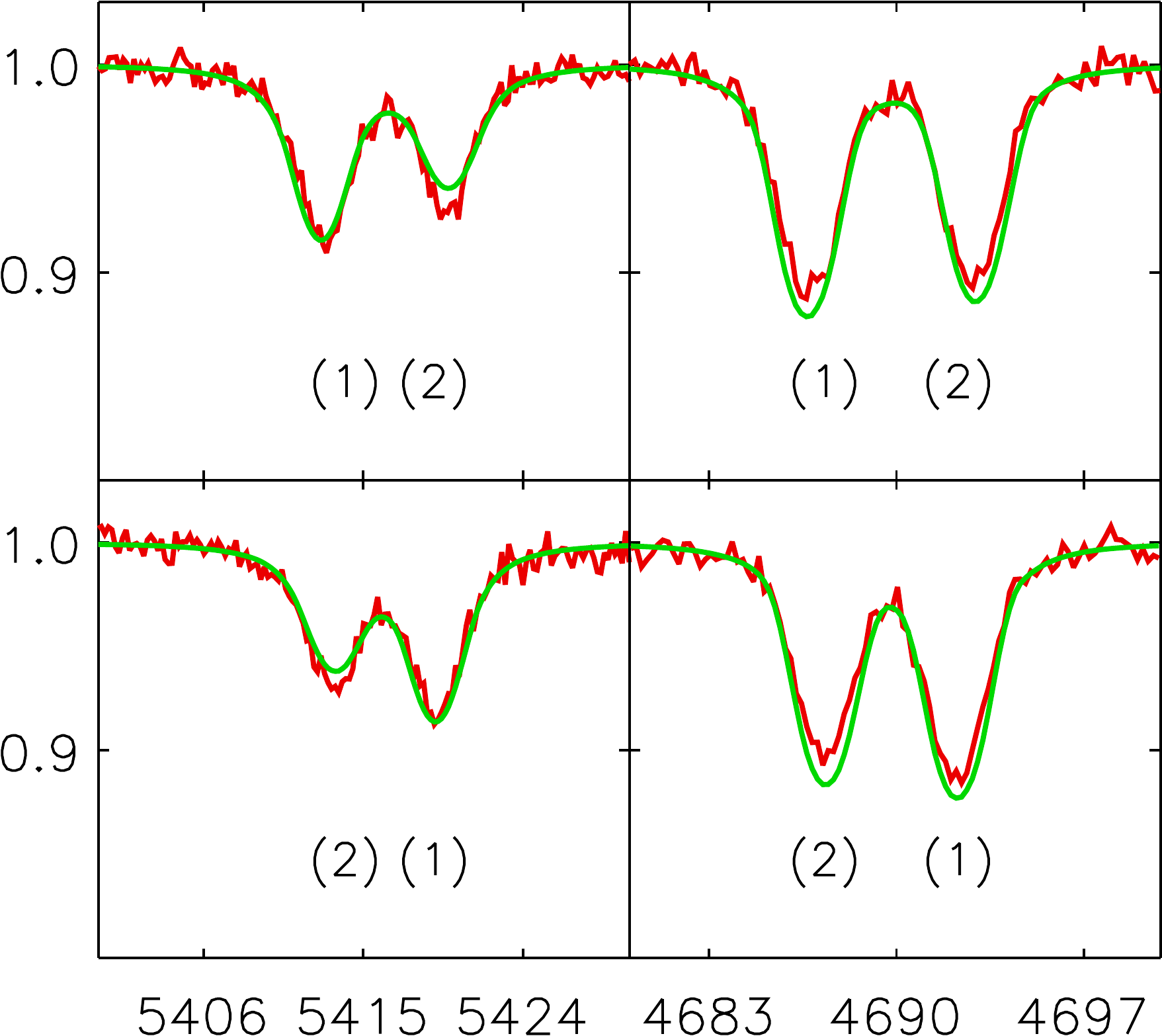} 
  \end{center}
  \caption{
Left figure: Fit of HeII 4542 MIKE spectrum. Right figure: fit of HeII 5411 (left) and HeII 4686 (right) with the UVES1 and UVES2 spectrum at the top and bottom, respectively.} \label{fig:He_4}
\end{figure*}

As previous NLTE studies of O-stars have shown \citep{Puls2005, Rivero2012a, Markova2018}, the fit of the helium lines requires the assumption of a depth-independent micro-turbulence velocity of 10 to 15 km/s in the NLTE line formation calculations and for the formal integral, which uses the helium NLTE occupation numbers to calculate the line profiles. We have, thus, adopted v$_{turb}$ = 10 and 15 km/s for our line profile calculations.
  
\subsection{Stellar mass-loss}

Stellar winds modify the outer atmospheric layers of hot massive stars through their outflow velocity field, which affects the formation of stronger spectral lines through Doppler shifts and a modified more extended mass density stratification, which causes additional wind emission (see \citealt{Kudritzki2000} for a review). It is, therefore, important to constrain the strengths of stellar wind line emission, which in the case of optical hydrogen and helium lines depends on the parameter $Q\,\sim\dot{M}\left(R_\star\,v_\infty\right)^{-3/2}$, where $\dot{M}$ and $v_\infty$ are the stellar wind mass-loss rate and the terminal velocity of the wind outflow, respectively (see \citealt{Puls2005, Holgado2018}). Terminal velocities are usually constrained through the observation of UV resonance lines of highly ionized metal lines, which show so-called P-Cygni profiles. Unfortunately, UV observations are not available for BLMC-02. But since the terminal velocities of winds of hot massive stars are very tightly correlated with the escape velocity from the stellar photosphere, we can use stellar gravity, radius and temperature to estimate $v_\infty$ using the formulae given in \cite{Kudritzki2000}. Assuming $v_\infty$ the mass-loss rates can then, in principle, be very precisely constrained through a fit of the hydrogen line H$_{\alpha}$ as described, for instance, by \cite{Puls1996} or \cite{Kudritzki2000}. The best fit for the two components of BLMC-02 is shown in Figure \ref{fig:halpha} for the case of R$_1$ = 8.85 R$_{\odot}$ and a primary effective temperature of T$_1$ = 35500K. The terminal velocities and mass loss rates adopted for this fit are $v_\infty$ = 2250 km/s and $\dot{M}$ = 2.0~10$^{-7}$ M$_{\odot}$/yr for the primary and  $v_\infty$ = 2380 km/s and $\dot{M}$ = 2.6~10$^{-8}$ M$_{\odot}$/yr for the secondary. The mass-loss rate for the primary is well constrained ($\pm$ 20 percent), because the absorption core of H$_{\alpha}$ is filled with a significant amount of wind emission. The mass-loss rate for the secondary, on the other hand, is very uncertain, because the effect of wind line emission is much smaller.  We note that the mass-loss rate of the secondary seems significantly smaller than the one of the primary indicating that it may belong to the class of 'weak-wind stars', which are frequently found among later spectral type main sequence O-stars \citep{Puls2008, Puls2009}. For models with different effective temperatures and different radii and gravities, terminal velocities are scaled with escape velocities and mass-loss rates are choosen so that the parameter $Q$ as defined above remains constant. In this way, fits of H$_{\alpha}$ are equally good for models with other stellar parameters.

 \begin{figure}[ht!]
  \begin{center}
    \includegraphics[scale=0.35]{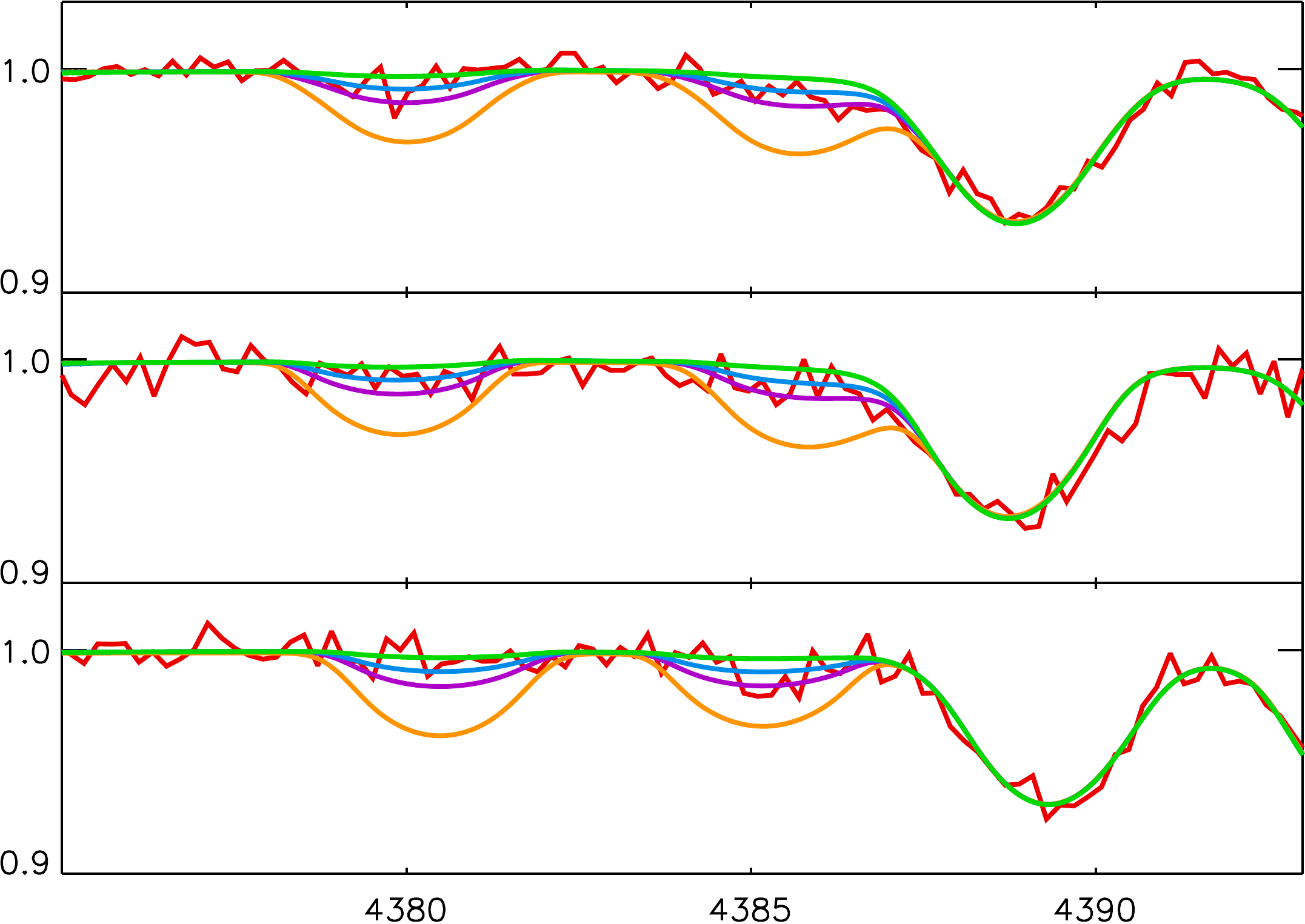}
    \includegraphics[scale=0.35]{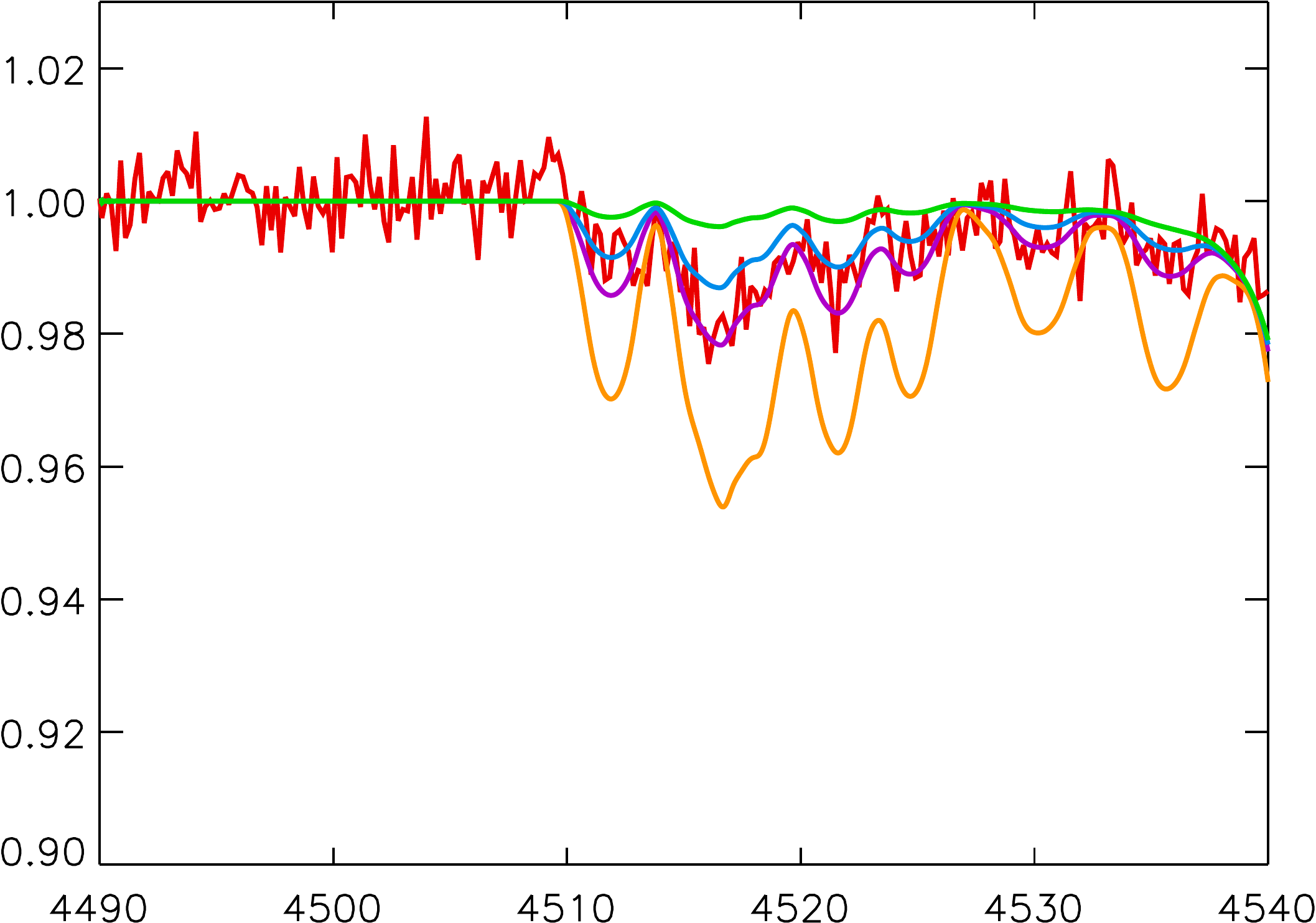} 
  \end{center}
  \caption{
Upper figure: Fit of NIII 4379 of primary and secondary with nitrogen abundances $\epsilon_{N}$ = 6.9 (green), 7.5 (blue), 7.8 (violet), and 8.5 (orange) of the MIKE (top), UVES1 (middle), and UVES2 (bottom) spectra. The line on the right is HeI 4387 from the primary star for MIKE and UVES1 spectrum, and from the secondary for the UVES2 spectrum. Lower figure: The nitrogen quartet lines around 4520\AA~of primary and secondary in the MIKE spectrum fitted with the same abundances as in the upper figure. } \label{fig:nitro_1}
\end{figure}

 \begin{figure}[ht!]
  \begin{center}
    \includegraphics[scale=0.35]{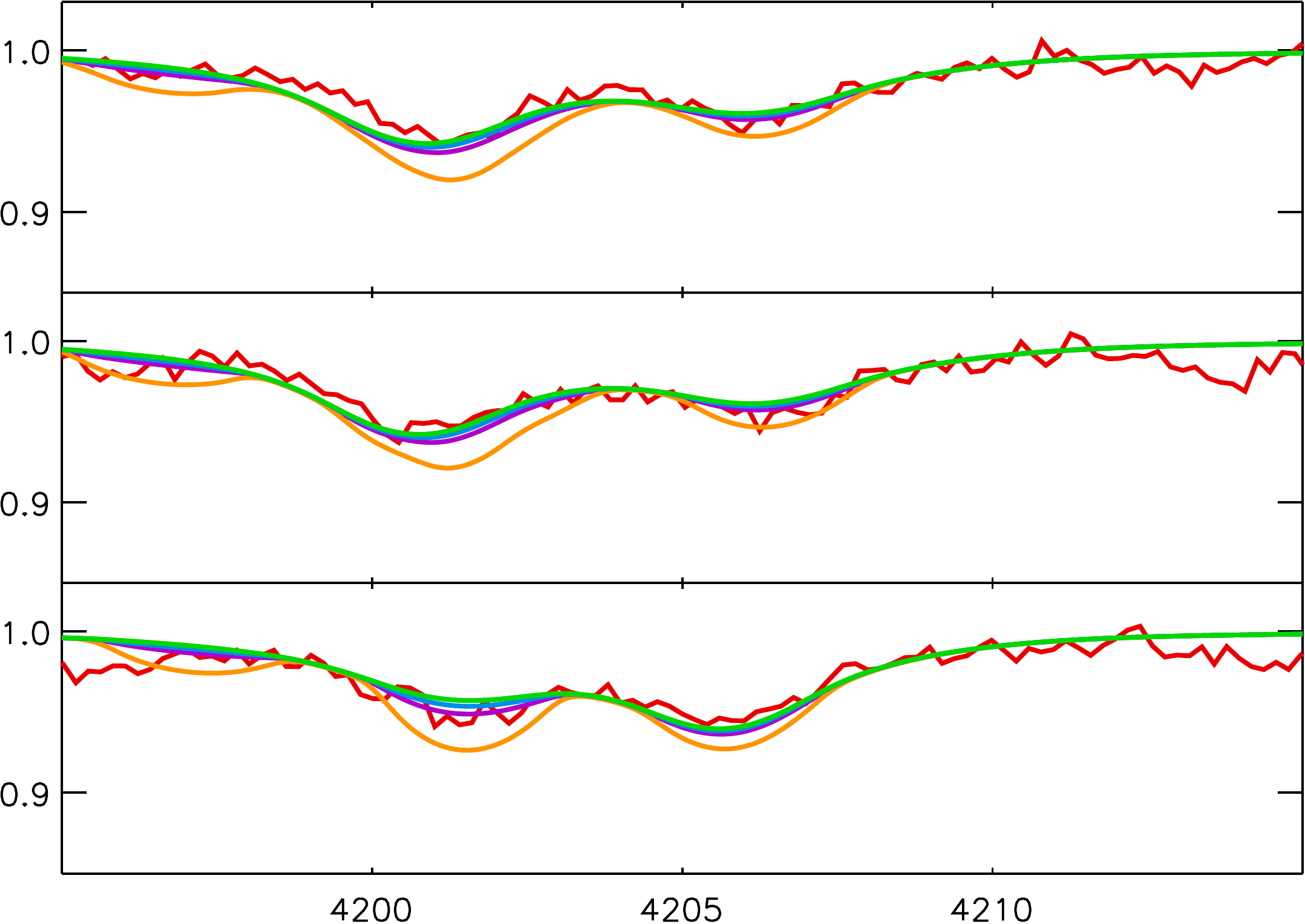}
    \includegraphics[scale=0.35]{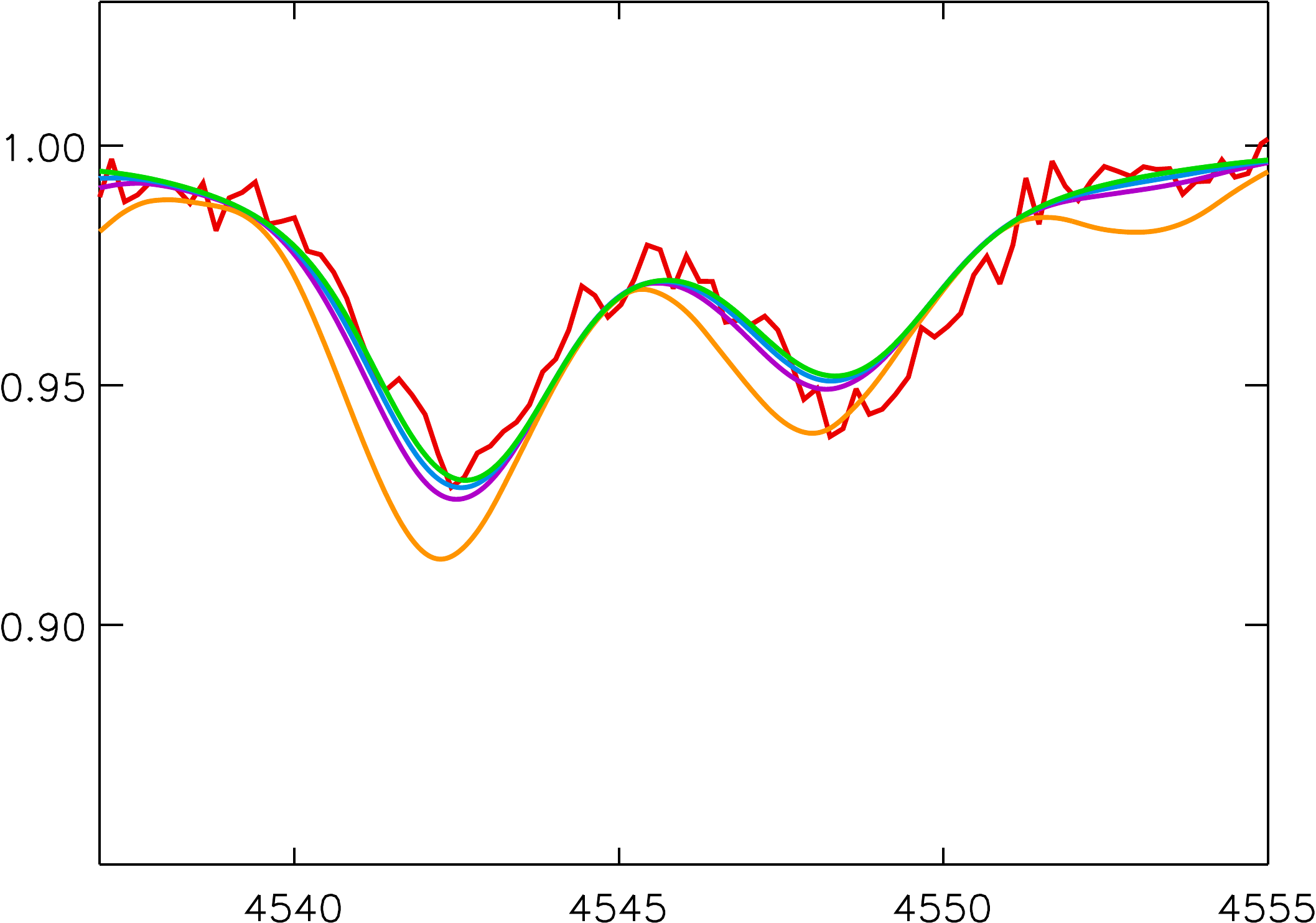} 
  \end{center}
  \caption{Same as Figure \ref{fig:nitro_1} but for the NIII blends with HeII 4200 (upper figure) and HeII 4542 (lower figure).} \label{fig:nitro_2}
 \end{figure}

For our line profile calculations including the effects of stellar winds we assume that the wind density structure is homogeneous. This is an approximation, because it is well known that the winds of hot stars are affected by so-called wind clumping (see review by \citealt{Puls2008}). In principle, spectral diagnostics can be used to constrain the amount of clumping and the related filling factors within the stellar wind outflow (see, for instance, \citealt{Puls2006} and \citealt{Kudritzki2006}). However, at the relatively low mass-loss rates that we encounter for BLMC-02 the effects of stellar wind clumping on the helium lines, which we use for the diagnostics of stellar temperature and radius, is of minor importance, as long as the stellar wind Q-parameters are chosen in an appropriate way so that the line with the strongest wind effect, H$_{\alpha}$, is well reproduced.

\subsection{Analysis technique}

For our analysis we use a two dimensional grid of atmosphere models with R$_1$ (and the corresponding values of R$_2$, F$_2$/F$_1$, log g$_1$, log g$_2$ as specified in Table \ref{tab:radtab}) and T$_1$ as free parameters, where T$_1$ ranges from 32000K to 38000K in steps of 100K. For each model we calculate normalized composite spectra accounting for the observed rotational velocities (105 km/s for each star) and the spectral resolution of the spectrographs by convolving with the corresponsung broadening profiles. As the profile fits are of good quality, we do not account for macro-turbulence velocities as an additional broadening mechanism.

We then select a set of HeI and HeII lines, for which we compare model and observed spectra by calculating $\chi^2$ - values. Figure \ref{fig:chi_lines} gives an example for one selected primary radius  and a helium abundance n(He)/N(H) = 0.09. For each primary radius R$_1$ and temperature T$_1$ we then calculate a total $\chi^2$(R$_1$, T$_1$) as sum of all the $\chi^2$-values of the individual helium lines. This will allow us to constrain temperature and radius of the components of the system as described in the next subsection.

\begin{figure*}[t]
  \begin{center}
    \includegraphics[scale=0.65]{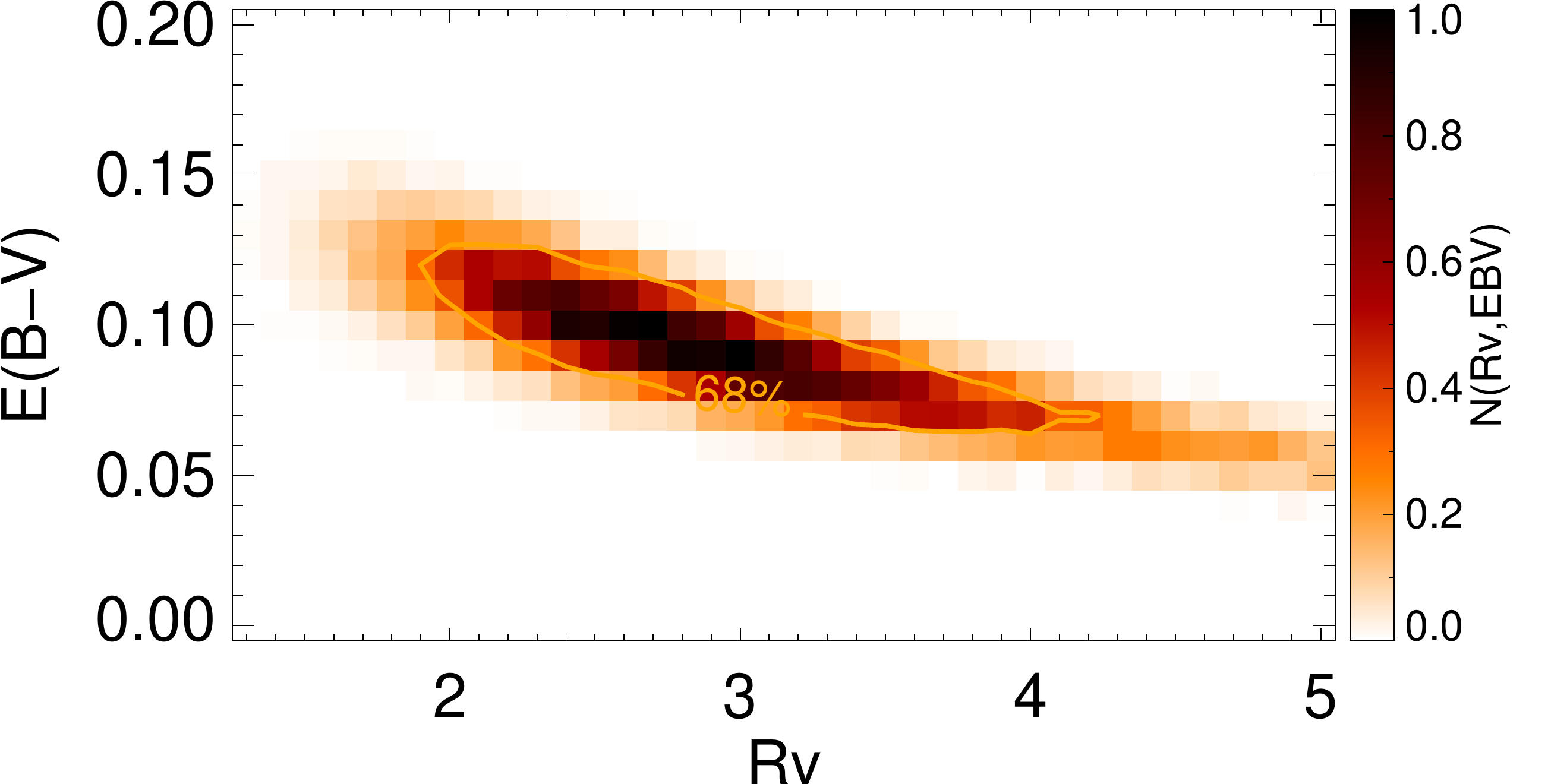}
  \end{center}
  \caption{Reddening and extinction of BLMC-02: Distribution of (R$_V$, E(B-V))-pairs obtained by the Monte Carlo fit of observed colors described in the text. The area enclosing 68\% of the distribution is indicated. The reddening law by \cite{Fitzpatrick1999} was used for this example. The fits with other reddening laws (see text) give similar results.} \label{fig:rvebv}
\end{figure*}

 \begin{figure*}[ht!]
  \begin{center}
    \includegraphics[scale=0.48]{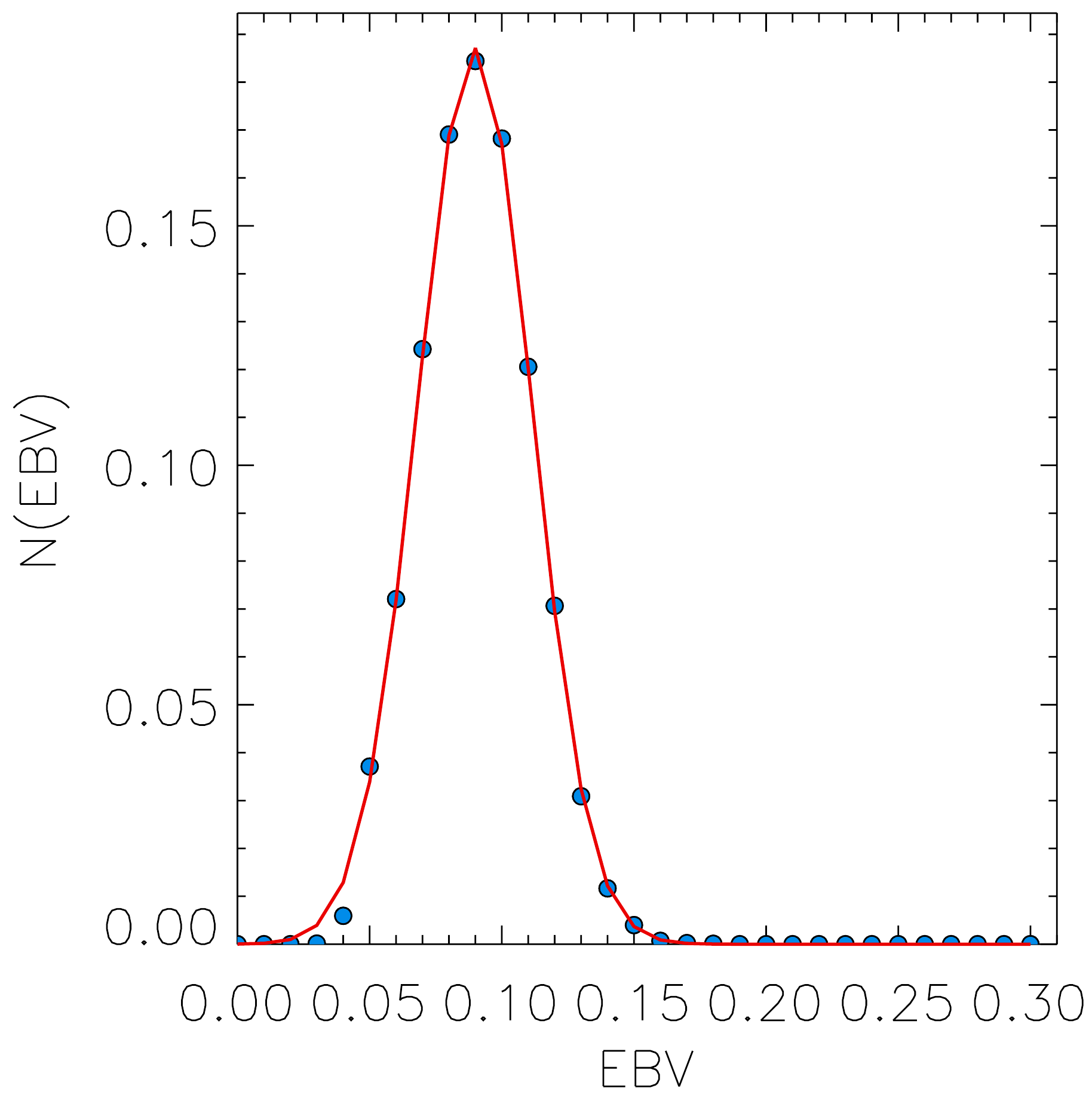}
    \includegraphics[scale=0.48]{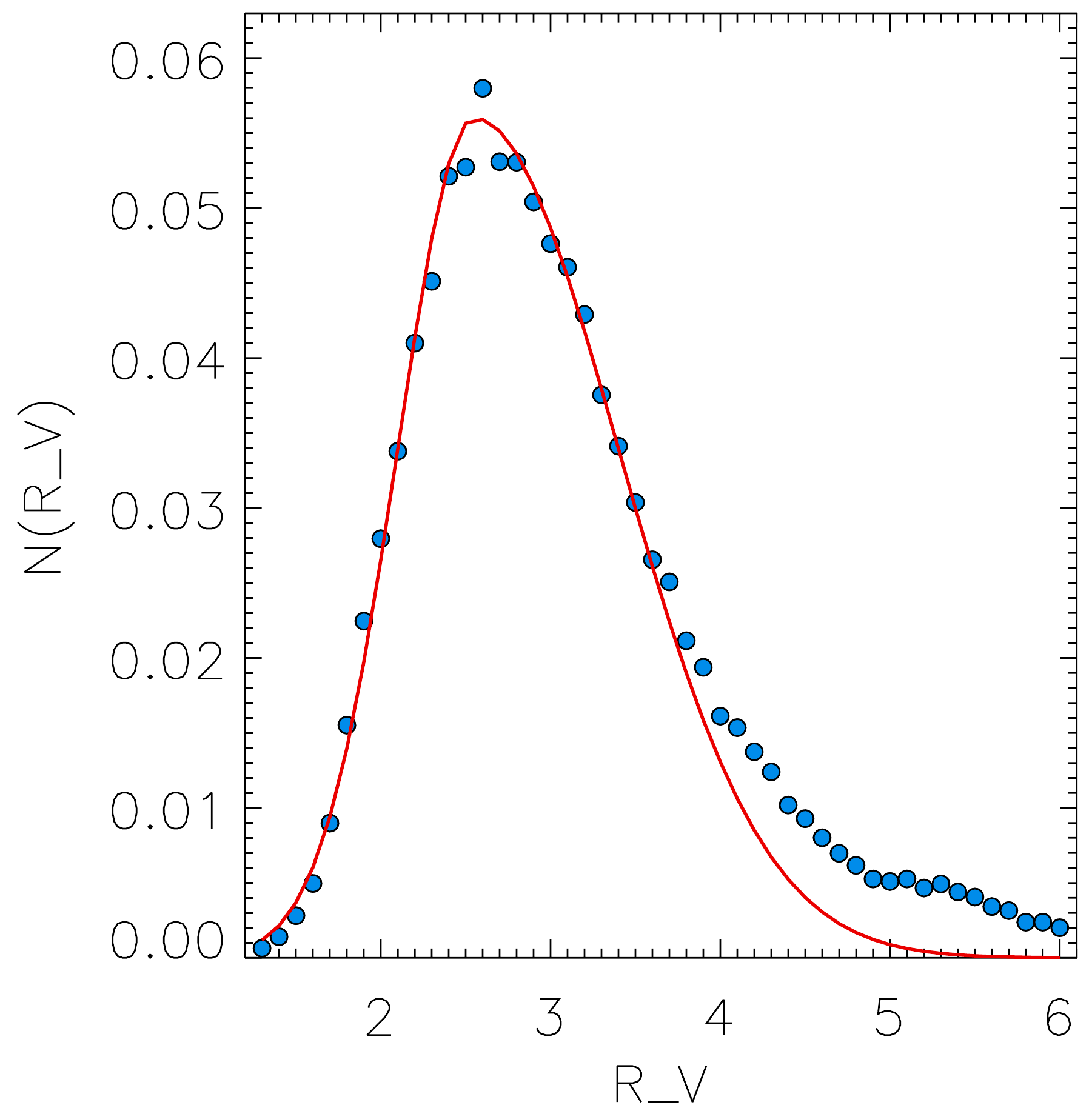}
    \includegraphics[scale=0.48]{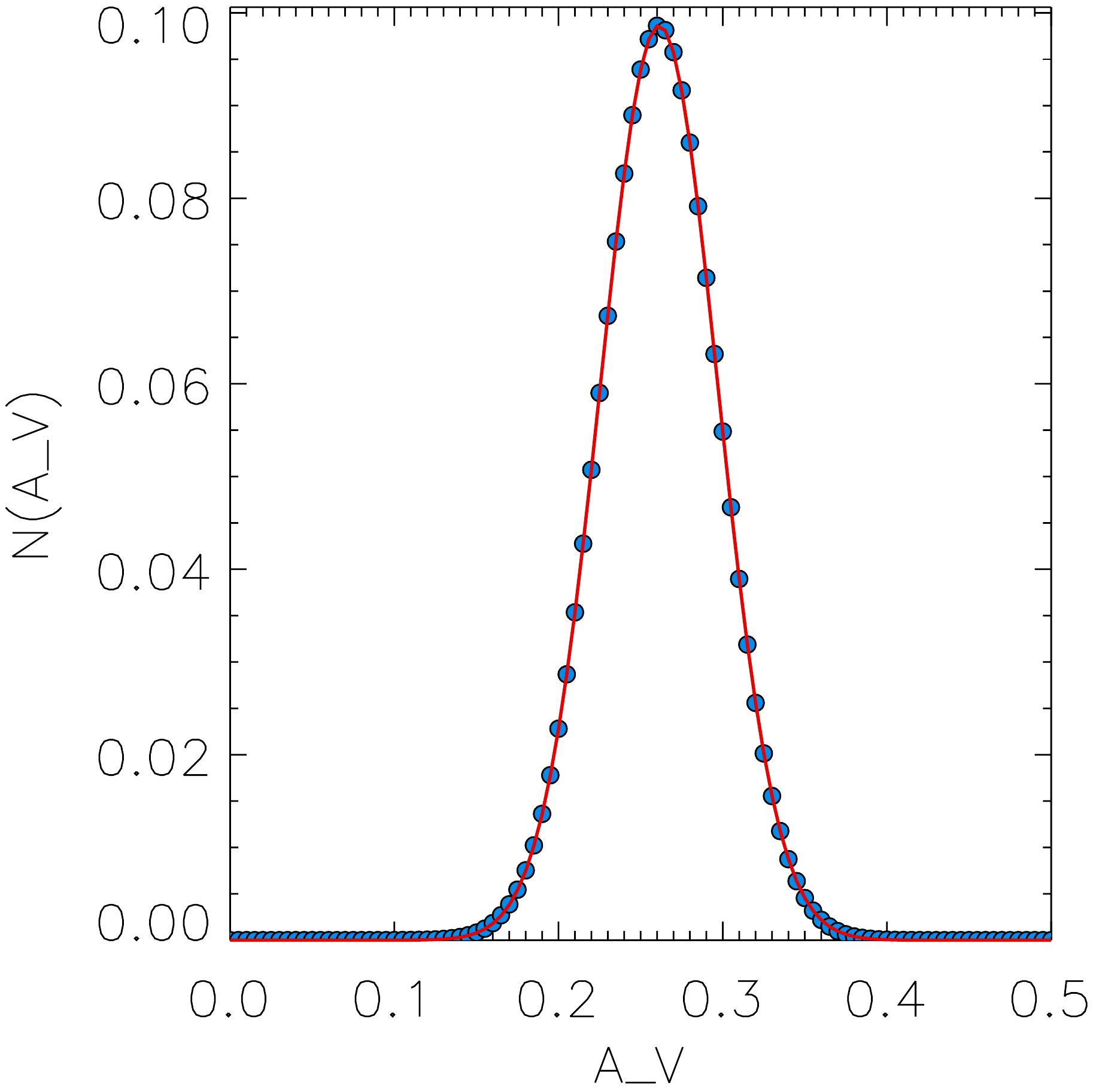}
    \includegraphics[scale=0.48]{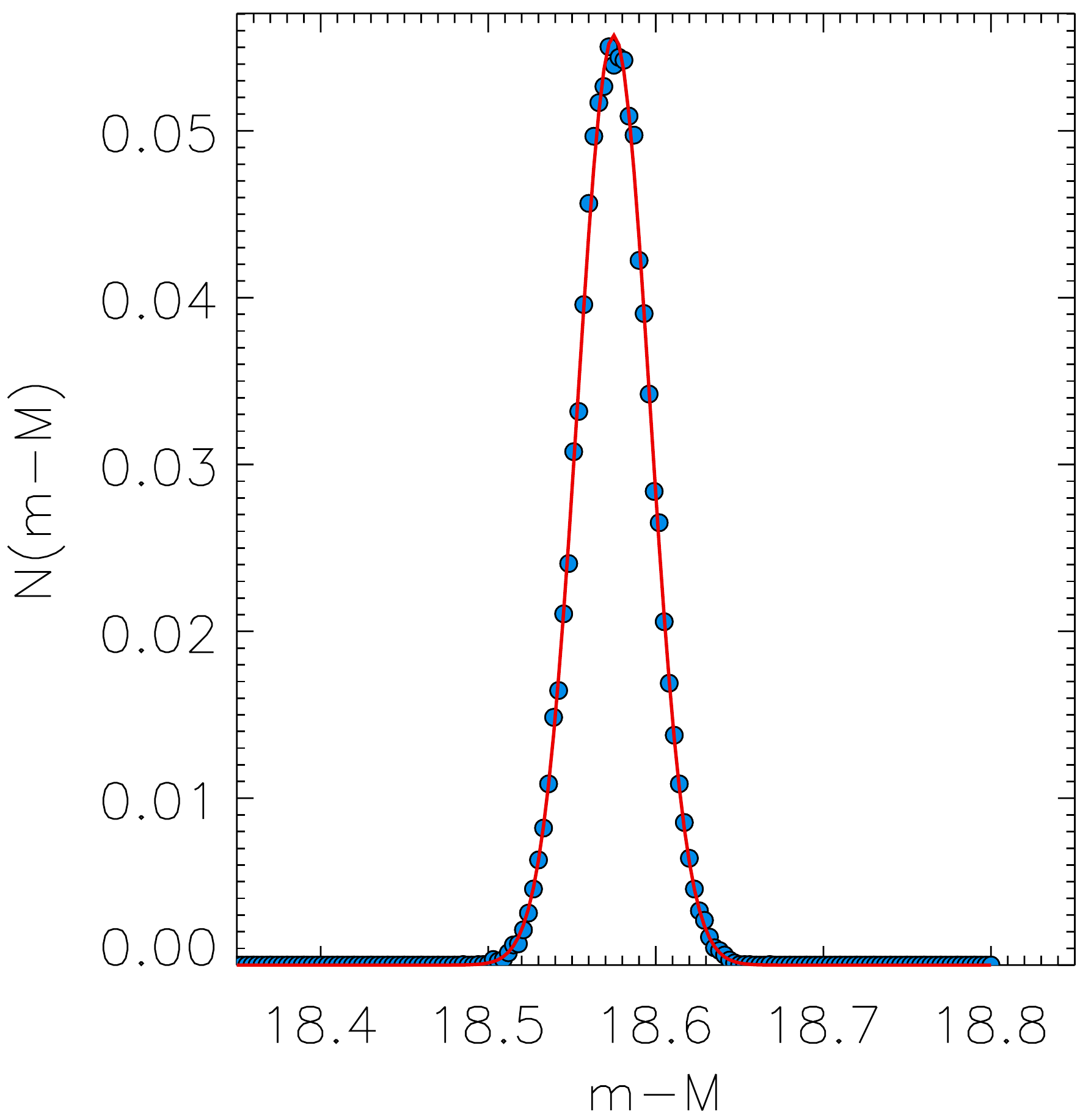}
  \end{center}
  \caption{Probability distributions (blue circles) and Gaussian fits (red) obtained from the Monte Carlo fit of observed colors shown in Figure \ref{fig:rvebv}. Top left: N(E(B-V)), top right: N(R$_V$), bottom left: N(A$_V$), bottom right: N(m-M).} \label{fig:probdm}
\end{figure*}
 
\begin{deluxetable*}{cccccccccc}
\tabletypesize{\small}
\tablewidth{0pt}
\tablenum{1}
\tablecolumns{10}
\tablecaption{Stellar radii, gravities and surface brightnesses ratios for the calculation of spectra} \label{tab:radtab}
\tablehead{
  \colhead{  } & \colhead{(1)} & \colhead{(2)} & \colhead{(3)} & \colhead{(4)} & \colhead{(5)} & \colhead{(6)} & \colhead{(7)} & \colhead{(8)} & \colhead{(9)}
}
\startdata
R$_1$ [R$_{\odot}$] & 9.150 & 9.100 & 9.050 & 9.000 & 8.950 &  8.900 & 8.850 & 8.800 & 8.750 \\
R$_2$ [R$_{\odot}$] & 7.590 & 7.660 & 7.730 & 7.795 & 7.860 &  7.920 & 7.980 & 8.040 & 8.100 \\
F$_2$/F$_1$        & 0.976 & 0.978 & 0.980 & 0.982 & 0.983 &  0.985 & 0.986 & 0.988 & 0.989 \\
log g$_1$ (cgs)    & 3.810 & 3.815 & 3.820 & 3.824 & 3.830 &  3.834 & 3.839 & 3.844 & 3.850 \\
log g$_2$ (cgs)    & 3.960 & 3.953 & 3.945 & 3.938 & 3.930 &  3.923 & 3.915 & 3.908 & 3.900 \\
\enddata
\vspace{-0.6cm}
\end{deluxetable*}

  \begin{figure}[ht!]
  \begin{center}
    \includegraphics[scale=0.50]{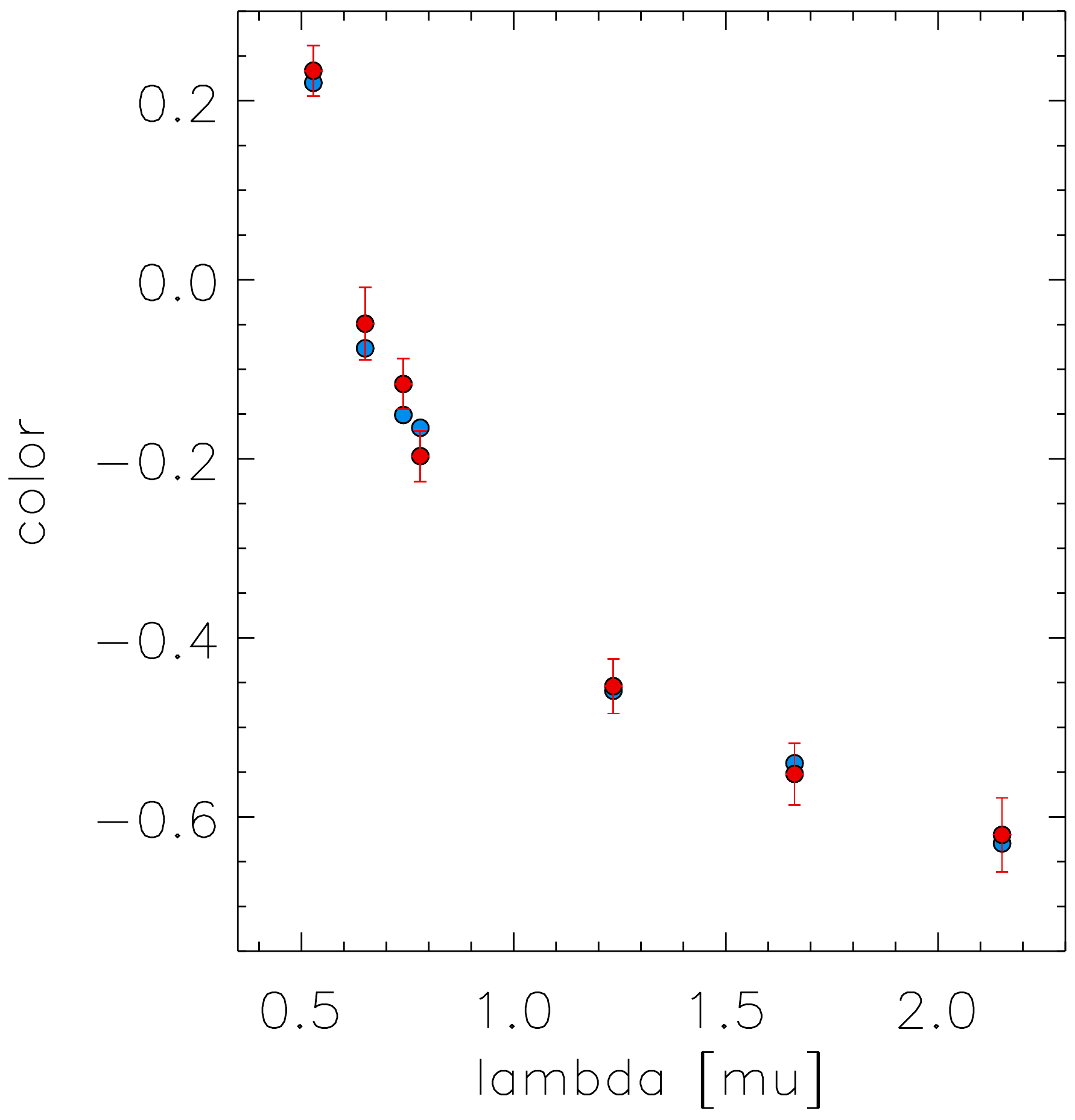}
  \end{center}
	  \caption{ Observed colors (V-G$_B$, V-R, V-G$_R$, V-I, V-J, V-H, V-K$_s$) of BLMC-02 (red) compared with model colors (blue) reddened with E(B-V) = 0.09 mag, R$_V$ = 2.9 and the reddening law by  \cite{Fitzpatrick1999}.  The colors are plotted as a function of the effective wavelength of their second filter passband.} \label{fig:colorfit}
\end{figure}

  \begin{figure}[ht!]
  \begin{center}
    \includegraphics[scale=0.50]{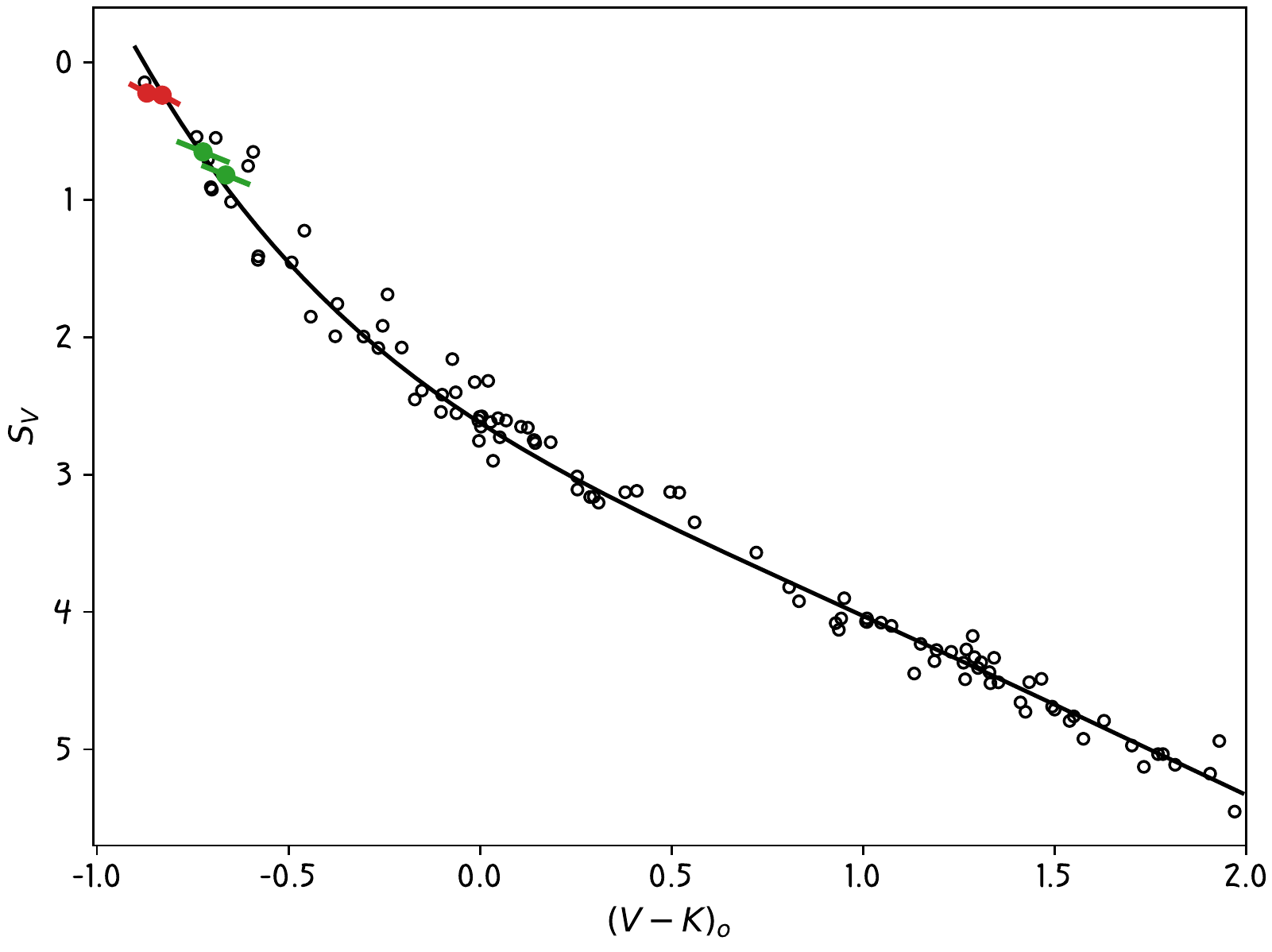}
  \end{center}
      \caption{Stellar surface brightness - color relationship. Interferometric measurements (small open circles) and the corresponding fit (solid line) by \cite{Challouf2014} are shown together with the values for BLMC-02 in red. The green points correspond to BLMC-01 which has been studied in Paper I.} \label{fig:surf}
\end{figure}

\subsection{Results of spectral analysis}

The calculation of $\chi^2$-values as described above has been carried for the whole range of helium abundances n(He)/n(H) between 0.08 and 0.15 and for micro-turbulences v$_{turb}$ = 10 and 15 km/s, respectively. The lowest $\chi^2$-values were obtained with v$_{turb}$ = 10 km/s for the primary and 15 km/s for the secondary. As for the helium abundance, the enhanced value 0.15 can be ruled out, because the $\chi^2$ minimum for the total of all HeII lines occurs at a significantly lower effective temperature than the minimum for the total of the HeI lines. The best solution is obtained for n(He)/n(H) = 0.09, where the $\chi^2$ minima of the neutral and ionized helium lines are at very similar effective temperature T$_1$ and the total $\chi^2$ of all the helium lines is the lowest. Figure \ref{fig:chisq} shows the isocontours $\Delta \chi^2$ = $\chi^2$(R$_1$, T$_ 1$) - $\chi^2_{min}$ in the (R$_1$, T$_1$)-plane, which encompass 68\% and 95\%, respectively, of the individual solutions obtained in extensive Monte Carlo simulations of the line fitting process. From the location of the 68\%-isocontour we read off the values T$_1$ = 35500 $\pm$ 200 K and R$_1$ = 8.85 $\pm$ 0.1 R$_{\odot}$ for the primary star. From Table \ref{tab:radtab} and our value of $\Delta$T = 1000K introduced above we then obtain for the secondary T$_2$ = 34500 $\pm$ 200 K and R$_2$ = 7.98 $\pm$ 0.06 R$_{\odot}$.

 \begin{figure*}[ht!]
  \begin{center}
    \includegraphics[scale=0.5]{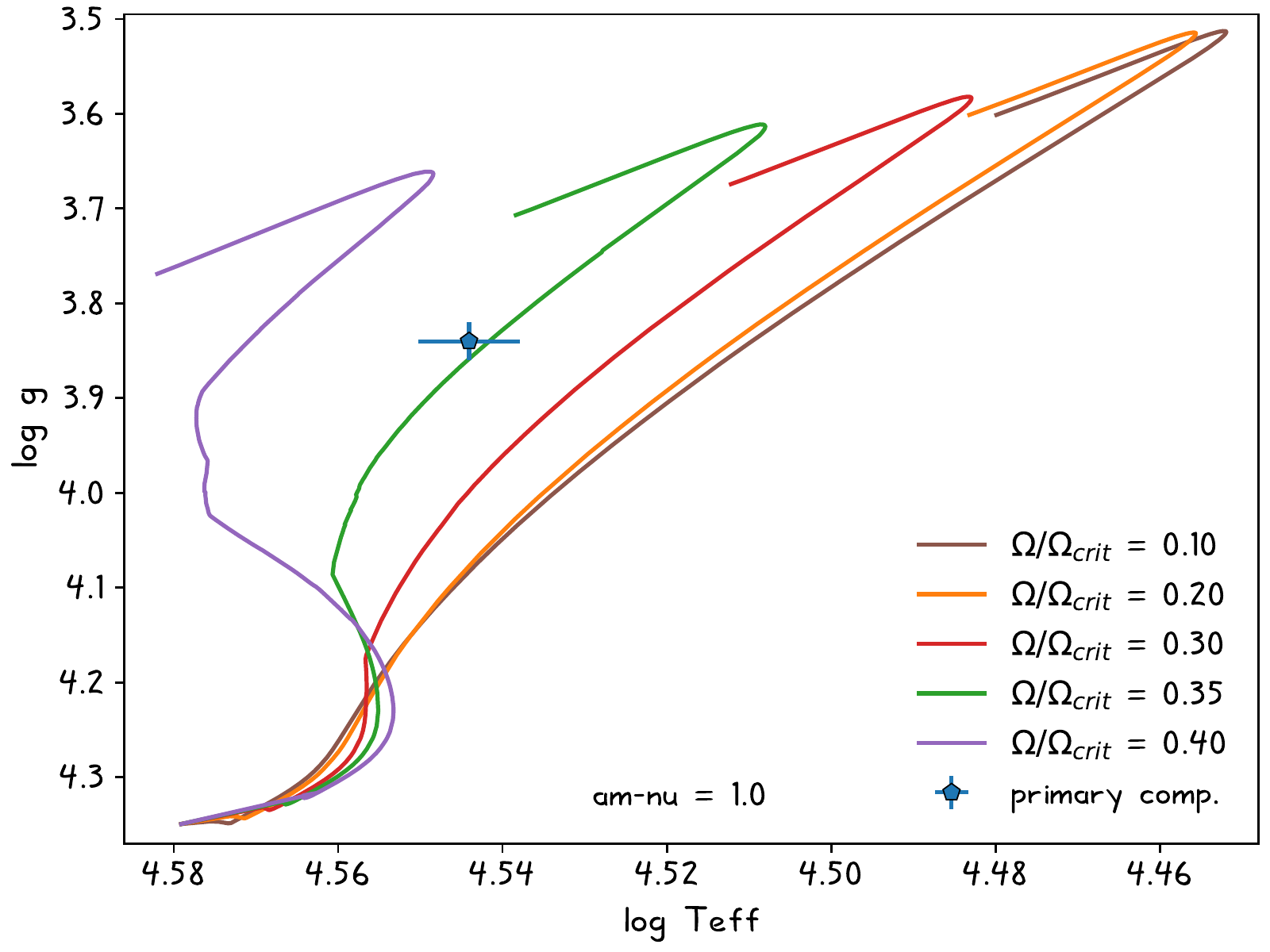}
    \includegraphics[scale=0.5]{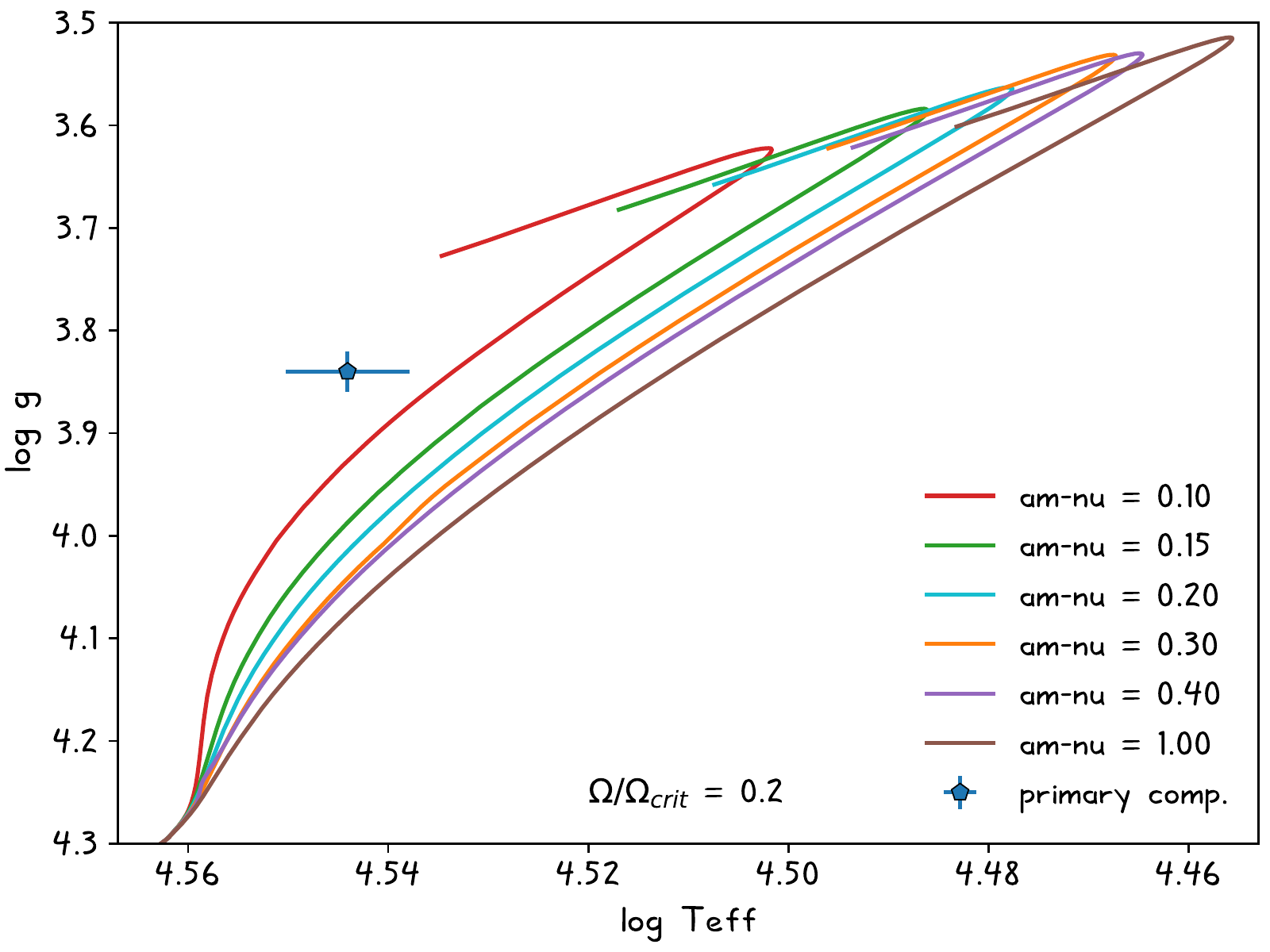}
  \end{center}
  \caption{(log g, log Teff)-diagram of stellar evolution. Left: The evolution of a star with an initial mass of M/M$_{\odot}$ = 20 and 
    different values of $\Omega$/$\Omega_{\rm crit}$.
   The \texttt{MESA} standard angular momentum transport coefficient is used for these tracks. Right: M/M$_{\odot}$ = 20, $\Omega$/$\Omega_{\rm crit}$ = 0.2 and angular momentum transport coefficient multiplied by a factor 0.1 (red), 0.15 (green), 0.2 (cyan), 0.3 (orange), 0.4 (purple), 1.0 (brown).The position of the primary star is indicated by the blue pentagon with error bars.} \label{fig:lgt1}
\end{figure*}

 \begin{figure}[ht!]
  \begin{center}
    \includegraphics[scale=0.5]{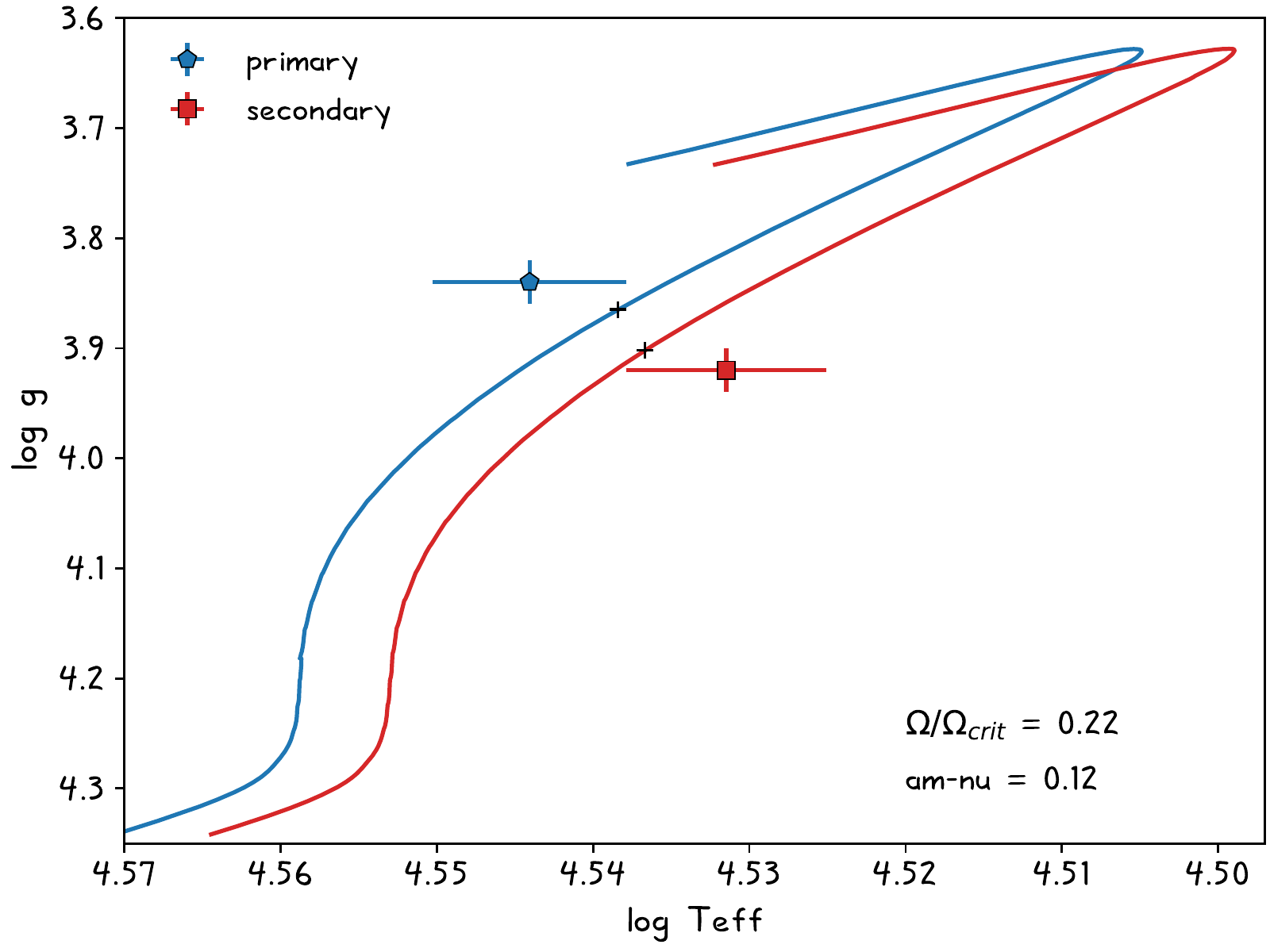}
  \end{center}
  \caption{(log g, log Teff)-diagram of stellar evolution. Evolutionary tracks for the primary and secondary component as described in the text. The positions of the primary and secondary are indicated by the blue pentagon and red square, respectively. The evolution time of 8.7 Myr is indicated as a cross on each track.} \label{fig:lgt2}
\end{figure}
 
\begin{deluxetable*}{ccccc}
\tabletypesize{\small}
\tablewidth{0pt}
\tablenum{2}
\tablecolumns{5}
\tablecaption{Reddening, extinction and distance modulus of BLMC-02 as obtained from the Gaussian fits described in the text.} \label{tab:exti}
\tablehead{
 \colhead{reddening} & \colhead{R$_V$} & \colhead{E(B-V)} & \colhead{A$_V$} & \colhead{m-M} \\
 \colhead{  } & \colhead{  }    & \colhead{mag}    & \colhead{mag}   & \colhead{mag} }
\startdata
ODON    & 2.60$^{+0.95}_{-0.62}$ & 0.090$\pm$0.021 & 0.26$\pm$0.04 & 18.56$\pm$0.04 \\
CARD    & 2.50$^{+1.02}_{-0.55}$ & 0.091$\pm$0.022 & 0.26$\pm$0.04 & 18.56$\pm$0.04 \\
MAIZ    & 2.60$^{+0.95}_{-0.50}$ & 0.089$\pm$0.020 & 0.26$\pm$0.04 & 18.56$\pm$0.04 \\
FITZ    & 2.55$^{+0.85}_{-0.45}$ & 0.090$\pm$0.022 & 0.26$\pm$0.04 & 18.57$\pm$0.04 \\
adopted & 2.89$^{+0.77}_{-0.77}$ & 0.090$\pm$0.022 & 0.26$\pm$0.04 & 18.56$\pm$0.04 \\
\enddata
\vspace{-0.6cm}
\end{deluxetable*}

It is important to check the result of the $\chi^2$-analysis  by a detailed comparison of observed helium line profiles with the stellar atmosphere model profiles calculated for the best model parameters. This is done in Figures \ref{fig:He_1} to \ref{fig:He_4} for the individual lines and the MIKE, UVES1 and UVES2 spectra. Note that for some lines we do not use all three sets of observed spectra, because of gaps in the wavelength coverage or inferior S/N ratio or obvious perturbations by cosmic rays etc. (see also Figure \ref{fig:chi_lines}).

\begin{deluxetable}{lcc}
\tabletypesize{\small}
\tablewidth{0pt}
\tablenum{3}
\tablecolumns{3}
\tablecaption{Stellar Parameters} \label{tab:stellarprop}
\tablehead{
  \colhead{  } & \colhead{primary} & \colhead{secondary}
}
\startdata
M/M$_{\odot}$    & 19.62$\pm$0.19 & 19.05$\pm$0.14 \\
R/R$_{\odot}$    & 8.85$\pm$0.10 & 7.98$\pm$0.12 \\
T$_{\rm eff}$ [K]    & 35000$\pm$500 & 34000$\pm$500 \\
log g [cgs]     & 3.84$\pm$0.02 & 3.92$\pm$0.02 \\
log L/L$_{\odot}$ & 5.024$\pm$0.03 & 4.884$\pm$0.03 \\
S$_V$ [mag] & 0.224$\pm$0.06 & 0.239$\pm$0.06 \\
(V-K$_S$)$_0$ [mag] & -0.87$\pm$0.04 & -0.83$\pm$0.04 \\
spectral type & O7.5 & O7.5 \\
$v_{turb}$ [km~s$^{-1}$]  & 10 & 15 \\
$v_{rot} \, \sin i$ [km~s$^{-1}$]  &  105 & 105 \\
$\dot{M}$  [M$_{\odot}$/yr]  & 2.0~10$^{-7}$  &  2.6~10$^{-8}$ \\
obital period [days] & \multicolumn{2}{c}{ 4.2707640 $\pm$ 3.3e-07} \\
semimajor axis [$R_\odot$] &   \multicolumn{2}{c}{ 37.46 $\pm$ 0.11} \\
inclination &  \multicolumn{2}{c}{ 86.14 $\pm$ 0.15  } \\
eccentricity &\multicolumn{2}{c}{ 0.0809  $\pm$ 0.0008 } \\
$\omega$ [rad] & \multicolumn{2}{c}{ 0.995 $\pm$ 0.009  } \\
$\dot{\omega}$ [rad/d] &  \multicolumn{2}{c}{ 0.000145 $\pm$ 2.7e-06 } \\
\enddata
\vspace{-0.6cm}
\end{deluxetable}

The fit of the HeI lines is generally very good, although we sometimes see small differences between the different observed spectra (compare, for instance, HeI 4471 MIKE versus UVES1, or HeI 5048 UVES1 versus UVES2 in Figure \ref{fig:He_2}), which we attribute to the data reduction or small stellar variability. The fits of the HeII lines are also good, but we note two systematic effects. First, for HeII 4686 the absorption depths of the model profiles for both components are a little too strong for the UVES1 and the UVES2 spectrum (and also for the MIKE spectrum, which is not shown in Figure \ref{fig:He_4}). We interprete this small discrepancy as a subtle effect of stellar wind clumping (see subsection 4.2). HeII 4686 is affected by stellar winds more strongly than the other optical helium lines and its line cores form in a region of the wind outflow, where double ionized helium has significantly recombined into HeII. As a result, the cores of this line react differently to deviations from a homogeneous flow when compared with H$_{\alpha}$, which we have used to constrain the wind mass-loss rate and wind density. In principle, we could use the observed discrepancy to constrain the stellar wind filling factor using the technique described by \cite{Kudritzki2006}, however, this is clearly beyond the scope of this paper, because the effects on the determination of stellar parameters are very small. Anyhow, in order to be on the safe side with the analysis we have not included HeII 4686 in the $\chi^2$ analysis. Second, for all the other HeII lines fitted the model profiles for the primary are a little too strong and a bit too weak for the secondary. This small discrepancy goes away, if we adopt a smaller temperature difference $\Delta$T between primary and secondary, for instance $\Delta$T = 0. However, the adopted $\Delta$T = 1000K at T$_1$ = 35500K leads to a model surface brightness ratio of secondary to primary V-band flux of F$_2$/F$_1$ = 0.982, wheres for $\Delta$T = 0 we obtain F$_2$/F$_1$ = 1.024 in clear disagreement with the observed value. We, thus, keep the value of $\Delta$T = 1000K. We note that reducing the value to $\Delta$T = 0 would have a small effect on the model SED for the composite system affecting the determination of distance modulus by $\Delta$(m-M) $\sim$ 0.02 mag only.

\subsection{HeII blends by nitrogen}

The model calculations for the hydrogen and helium lines of our analysis do not include potential blends by metal lines, as the contribution of metal lines to the optical spectra of O-stars is usually weak, in particular, at the lower metallicity of the Magellanic Clouds. However, it is well known that nitrogen lines of NIII can affect the HeII lines HeII 4200 and 4541, most importantly in cases, when the atmospheres of O-stars are enriched through rotational mixing with material from the stellar interior, which has undergone the CN-cycle. We have, therefore, carried out detailed non-LTE line formation calculations using the the atomic models and methods described by \cite{Rivero2012a, Rivero2012b} for the final atmospheric model of BLMC-02 to constrain the nitrogen abundance and the effects of NIII blends on the HeII lines. We have adopted four values of the nitrogen abundance $\epsilon_{N}$ = 6.9, 7.5, 7.8, 8.5 defined as $\epsilon_{N}$ = log n(N)/n(H) +12. Figure \ref{fig:nitro_1} shows the result for the prominent nitrogen line NIII 4379. In agreement with normal helium abundance n(He)/N(H) = 0.09 we can safely exclude a strong enhancement of nitrogen abundance through the CN-cycle and estimate a value of $\epsilon_{N}$ = 7.5 or, at most, 7.8. This result is confirmed by the comparison with the NIII quartet lines at 4520\AA~shown in the same figure. For this nitrogen abundance the effects of nitrogen blending of the HeII lines are very small, as Figure \ref{fig:nitro_2} confirms. The nitrogen abundance obtained indicates a factor four (at most eight) increase through rotational mixing relative to the LMC standard value $\epsilon_{N}$ = 6.9  for massive early type stars (see \citealt{Hunter2008}).

\subsection{Systematic effect through turbulence pressure}

As described above, the fit of the helium lines required the assumption of a micro-turbulence of v$_{\rm turb}$ = 10 and 15 km/s for the primary and secondary, respectively. However, as for instance discussed by \cite{Markova2018}, micro-turbulence will lead to a turbulence pressure term in the stellar atmosphere momentum equation, which can be approximated by P$_{\rm turb}$= x$\rho$v$_{turb}^2$, where $\rho$ is the mass density and x is a factor of the order of 0.5 \citep{Josselin2007}. The gradient of turbulent pressure leads to an outward acceleration which effectively reduces the inward gravitational acceleration acting on a volume element in the O-star atmospheres. In the subsonic deeper layers of the stellar atmosphere, where the helium lines used for our spectral diagnostic are formed, this reduction can be straightforwardly estimated to be $\Delta$log g = -log(1+xv$_{\rm turb}^2$/v$_{\rm sound}^2$), where v$_{\rm sound}$ is the sound speed. In the  \texttt{FASTWIND}  code (and also in most of the other NLTE model atmosphere codes available), turbulence pressure is not included when the atmospheric density stratification is calculated from the momentum equation. Thus, when we fit the observed helium lines as described above with model atmospheres adopting the gravities obtained from the analysis of radial velocity and light curve, we use atmospheric models with effective gravities which are too large because of the neglect of turbulence pressure. We can account for this effect by using  \texttt{FASTWIND}  models, for which the gravities are corrected by $\Delta$log g, which is -0.04 dex for the primary and -0.09 dex for the secondary. From our available calculated grid of  \texttt{FASTWIND}  models we know that the use of models with these slightly lower gravities leads to a fit of the helium ionization equilibrium at effective temperatures 500K cooler than in the case where the effect of turbulence pressure is neglected (see also \citealt{Puls2005}, \citealt{Kudritzki2006}). We therefore adopt T$_1$ = 35000 $\pm$ 500K for the primary star and T$_1$ = 34000 $\pm$ 500K for the secondary.

We note that an analogous effect on the effective gravity is caused by stellar rotation and the centrifugal acceleration g$_{\rm centri}$ = v$_{\rm rot}^2$/R (see, for instance, \citealt{Herrero1992}). However, in the case of our objects with the rotational velocities, radii and gravities observed and accounting for an average of the rotation effect over the unresolved total stellar surface the influence on the gravity can be neglected.

The stellar parameters for both components of the system are summarized in Table \ref{tab:stellarprop}.

\section{Interstellar reddening and distance}

The estimate of interstellar reddening and extinction is crucial for the determination of surface brightness, intrinsic colors and distance. We proceed in a similar way as described in \cite{Urbaneja2017}. We use the composite SED of our final atmosphere models of the primary and secondary and redden this SED by assuming a grid of E(B-V) and R$_V$ = A$_V$/E(B-V) values in a range from 0.0 to 0.3 mag and 1.0 to 7.0, respectively. This is done by converting the filter passband values of E(B-V) and R$_V$ into monochromatic values E(4405-5495) and R$_{5495}$ = A$_{5495}$/E(4405-5495) and then applying the reddening and extinction of four different monochromatic reddening laws, \cite{Cardelli1989} (CARD), \cite{Fitzpatrick1999} (FITZ), \cite{Maiz2014,Maiz2017} (MAIZ) and \cite{Odonnell1994} (ODON). The reddened SEDs are then used to calculate reddened colors (G$_B$-V), (V-G$_R$), (V-R), (V-I), (V-J), (V-H), (V-K$_s$) in the same passbands as observed (see section 2) and a $\chi^2$-value is calculated from the difference between calculated and observed colors. This provides the best pair of E(B-V) and R$_V$ at the minimum of $\chi^2$. By carrying out a Monte Carlo simulation, where we modify observed colors with randomly drawn corrections assuming a Gaussian distribution of uncertainties corresponding to the observed errors given in section 2, we then construct probability distribution functions N(E(B-V)), N(R$_V$), N(A$_V$) and N(m-M) for reddening, the ratio of extinction to reddening, extinction and distance modulus, respectively.

  Figure \ref{fig:rvebv} shows the distribution of solutions N(R$_V$, E(B-V)) in the two-dimensional plane of R$_V$ and E(B-V) of a simulation with 30,000 runs and assuming a FITZ reddening law. The curve surrounding the area containing 68\% of the solutions is also shown and allows an assessment of the uncertainties of R$_V$ and E(B-V). The two-dimensional distribution can be marginalized to derive one-dimensional probability distributions N(R$_V$) and N(E(B-V)), which are displayed in Figure \ref{fig:probdm} together with their Gaussian fits. As already found in \cite{Urbaneja2017}, N(R$_V$) requires a two-component fit with two different standard deviations. The Gauss-fit is poor but still gives an idea of the uncertainty in R$_V$. Note that because of assymetric nature of N(R$_V$) the mean value $<R_V>$ is larger than the value for the maximum of the distribution.

For each pair (R$_V$,E(B-V)) we can de-redden the observed Monte-Carlo simulated magnitudes in each passband then use the model atmosphere magnitudes of the final composite model SED to calculate a distance modulus averaged over all passbands.  The distribution N(R$_V$, E(B-V)) then allows to construct the probability distribution N(m-M) for the distance modulus given in Figure \ref{fig:probdm} together with its Gaussian fit. The standard deviation of this contribution is 0.03 mag and much smaller than expected at first glance from the large uncertainty of R$_V$. The reason for this, as already pointed out in \cite{Urbaneja2017}, is the fact that the values for R$_V$ and E(B-V) found from the SED fitting are not independent but are instead anti-correlated, which limits the uncertainty range of extinction. This can be demonstrated by calculating A$_V$ = R$_V$E(B-V) for each point in Figure \ref{fig:rvebv} and then constructing the probability distribution function N(A$_V$) for visual extinction. This distribution together with the Gaussian fit is also given in Figure \ref{fig:probdm}.

Table \ref{tab:exti} summarizes the results for the different reddening laws. As an average over the four reddening laws we adopt E(B-V) = 0.09 $\pm$ 0.02 mag, A$_V$ = 0.26 $\pm$ 0.04 mag and a distance modulus m-M = 18.56 $\pm$ 0.04 mag. The uncertainty of the distance modulus is slightly larger than the $\sigma$-value of the Gaussian fit in Figure \ref{fig:probdm}, because we have added the effects of temperature and radius uncertainty in quadrature. The adopted value of R$_V$ in Table \ref{tab:exti} is the mean value $<R_V>$ = 2.89 as calculated from the averages of E(B-V) and A$_V$.
As discussed above it is slightly larger than the maxima of the distributions N(R$_V$) for the individual reddening laws, which are also given in Table \ref{tab:exti}.

For the calculation of our synthetic photometry we used the V, R, I filter functions and zero points by \cite{Bessell2012}. The 2MASS filter functions for J, H, and K$_s$ where downloaded from the 2MASS website and the zero points by \cite{Cohen2003} were used. For both sets of filters we checked the zero points by calculating Vega passband fluxes based on the HST Vega  stis005 spectrum (see \citealt{Bessell2012} for reference). We found excellent agreement with \cite{Bessell2012} and \cite{Cohen2003}. The Gaia G$_B$ and G$_R$ filter function (see \citealt{Evans2018}) were obtained from the Gaia website. To be consistent with our calculations for the other passbands we then determined zero points for these filters again by using  HST Vega  stis005 spectrum. Figure \ref{fig:colorfit} shows an example for the comparison between observed and calculated model atmosphere colors.

\section{The distance modulus to the LMC from BLMC-02}

As already discussed in Paper I, BLMC-02 is located slightly away from the center of the LMC and the line of nodes. Applying the geometric model for the LMC by \cite{vanderMarel2014} we estimate a difference in distance modulus $\Delta$(m-M) = 0.031 $\pm$ 0.009 mag between BLMC-02 and the center of the LMC. If we apply this correction to the distance modulus of BLMC-02 obtained in the previous section, we obtain a distance modulus to the center of the LMC of (m-M)$_{\rm LMC}^{\rm BLMC-02}$ = 18.53 $\pm$ 0.04 mag. This is 0.05 mag larger than he most recent, very accurate distance modulus to the LMC of (m-M)$_{\rm LMC}$ = 18.477 $\pm$ 0.026 mag measured by \cite{Pietrzynski2019} but agrees within the margin of the errors. The analysis of additional well detached systems of spectral type O using the same method as applied here is needed to  investigate whether the difference is random or points to a systematic effect of the model atmosphere analysis.

\section{Surface brightness and intrinsic (V-K$_s$) color}

With the determination of interstellar reddening and extinction and the constraints on the stellar radii, we can use the known distance to the LMC to measure the surface brightnesses of the components of the binary system. Following \cite{DiBenedetto2005} and \cite{Pietrzynski2019} we obtain the surface brightness S$_V$ in the V-band via

\begin{equation}
\theta = 9.2984{R \over d}
\end{equation}
and
\begin{equation}
S_V = m_V^0 + 5log\theta
\end{equation}

R is the stellar radius in units of the solar radius, d the distance in parsec and $\theta$ the angular diameter in milli-arsec. m$_V^0$ is the de-reddened stellar magnitude in the V-band.

The values determined in this way for the primary and secondary are given in Table \ref{tab:stellarprop} together with the de-reddened (V-K$s$) colors. Note that for the distance we have used (m-M)$_{\rm LMC}$ = 18.477 $\pm$ 0.026 mag \citep{Pietrzynski2019} but applied the correction $\Delta$(m-M) = 0.031 $\pm$ 0.009 mag accounting for the location of BLMC-02 in the LMC as discussed in the previous section. Figure \ref{fig:surf} shows the observed stellar surface brightness - color relationship by \cite{Challouf2014} and compares with the results of this work and of Paper I. Within the uncertainties we find agreement with Challouf et al. fit curve.

\section{Stellar evolution}

The well constrained stellar parameters of both components of BLMC-02 provide an opportunity for a comparison with the prediction by stellar evolution theory.
Please note, that in the following analysis we assume a single star evolution, as the stars had not yet time to interact strongly with only a minor tidal interaction taking place. We started by comparing our results with evolutionary tracks from \cite{Brott2011}, \cite{Choi2016} and also with Geneva models \citep{Ekstrom2012,Georgy2013}, however we could not find an agreement with our measurements. For the masses of the system components, all of these models predict too low temperatures that do not correspond with the line features observed in the spectra. To check if it is possible to obtain an agreement we have eventually used the \texttt{MESA} stellar evolution code version 11701 \citep{Paxton2011, Paxton2013, Paxton2015, Paxton2018, Paxton2019}. \texttt{MESA} allows us to calculate stellar evolution tracks specifically tailored to match the observed masses and then to check, whether stellar evolution also reproduces the other stellar parameters, such as effective temperature, radii, gravity and luminosity. We adopt LMC metallicity, a mixing length parameter $\alpha_{\rm MLT}$ = 1.6 and step function overshooting with an overshooting parameter $\alpha_{\rm OV}$ = 0.1. For the effects of mass-loss the \texttt{MESA} module applying the results obtained by \cite{Vink1999, Vink2000, Vink2001} is used. We also account for rotational enhancement of mass-loss. Eddington-Sweet circulation dynamical and secular shear instability, and the Goldreich-Schubert-Fricke instability \citep{Heger2000} that are induced by rotation and result in mixing are taken into account.
Semiconvection and the effects of internal magnetic fields based on the Spruit-Tayler dynamo \citep{Spruit2002} mechanism are also included.

Figure \ref{fig:lgt1} (left) shows a (log g, log T$_{\rm eff}$)-diagram with a sequence of evolutionary tracks for a massive star with an initial mass of 20 solar masses and different ratios of the angular rotational velocity to the critical rotational velocity $\Omega$/$\Omega_{\rm crit}$.
The tracks were calculated to the point where the core hydrogen abundance drops to zero, i.e. to the end of the main sequence.
We have chosen the (log g, log T$_{\rm eff}$)-diagram for this comparison, because the observed gravities and temperatures are very precisely constrained. The HR diagram gives a very similar result but is not shown for the sake of brevity. As can be seen, whether the tracks match the location of the primary star in this diagram depends crucially on the choice of the initial rotational velocity. 

It is possible to use calculations of the same type as shown in Figure \ref{fig:lgt1} (left) and to tune the initial mass and the initial value of  $\Omega$/$\Omega_{\rm crit}$ to obtain evolutionary tracks which come close to the stellar parameters (mass, effective temperature, radius, gravity, luminosity) of the primary and secondary star. However, these solutions predict rotational velocities of the order of 250 km/s, which are much higher than the observed values of 105 km/s. Thus, a mechanism is needed to reduce the outer rotational velocities, until the observed stage of the components of BLMC-02 is reached. The \texttt{MESA} package offers this possibility by multiplying the angular momentum transport coefficient by a factor am-nu, where am-nu=1.0 is the standard option and smaller values reduce the angular momentum transport from the interior to outer layers and thus lead to smaller rotational velocities. The right of Figure \ref{fig:lgt1} displays a sequence of models, where $\Omega$/$\Omega_{\rm crit}$ is fixed to 0.2 but am-nu is varied between 0.1 to 1.0. Obviously, the choice of the efficiency of the angular momentum transport has also a significant effect on the evolutionary tracks.

In Figure \ref{fig:lgt2} we combine the dependence on intial rotational velocity and angular momentum transport efficiency and calculate evolutionary tracks for the primary and secondary, which come close to the observed stellar properties. We assume $\Omega$/$\Omega_{\rm crit}$ = 0.22 and am-nu = 0.12 and initital masses of 19.99 and 19.30 M$_{\odot}$ for the primary and secondary, respectively.
Based on these calculations the age of the system is about 8.7 Myr as indicated in the figure. For this age the values of gravities and T$_{\rm eff}$ of the two stars agree within about 1-$\sigma$ with the observations, as do the calculated masses and luminosities.
The calculated rotational velocities (130 km/s for the primary and secondary, respectively) are marginally too high but reasonably close to the observations. 
The tracks predict a slightly enhanced helium abundance due to rotationally mixing of N(He)/N(H) = 0.12 (after starting at the ZAMS with 0.08), which mildly disagrees with the observed value of 0.09. The nitrogen enhancement relative to the starting value on the ZAMS is $\Delta \epsilon_N$ = 0.9 marginally in agreement with the constraints on the observed nitrogen abundance discussed in secion 4.5.

We note that the major reason for adopting a significant reduction of angular momentum transport efficiency combined with a relatively high initial rotational velocity is the requirement that the evolutionary tracks need to reproduce the high observed stellar effective temperatures. As is clear from Figure \ref{fig:lgt1} (left), evolutionary tracks with similar $\Omega$/$\Omega_{\rm crit}$ but with am-nu = 1 reach the observed gravities at temperatures 3500K cooler than obtained from the spectral analysis, which is outside the margin of our errors. Model spectra calculated at these lower temperatures show ionized helium lines, which are clearly too weak, and neutral helium lines, which are slightly too strong. Moreover, the effective temperatures obtained by our analysis agree with the spectral type O7.5 (defined by the observed relative strength of neutral and ionized helum lines) as confirmed by comprehensive independent work analyzing the spectra of individual O-stars (see, for instance, \citealt{Martins2015}). Thus, the only way to find agreement with stellar evolution is the assumption of reduced angular momentum transport. 

Although this assumption sounds strong, we would like to point out that a similar conclusion was reached by \citet{Groh2019} for low-metallicity stars. These authors compared MESA models from \citet{Choi2016} with their Geneva models and found that in the MESA models the angular momentum transport is stronger, with the reason being a different treatment of the meridional currents. \citet{Groh2019} solve an advective equation for the angular momentum transport, while the diffusive euation is used in MESA.
 
\section{Discussion and Future Work}

The work presented here provided three major results, the determination of a new independent distance to the LMC, a measurement of surface brightness for two early type stars and a test of stellar evolution.

The LMC distance modulus of 18.53 $\pm$ 0.04 mag obtained from our spectroscopic analysis is 0.05 mag larger than the precision distance of (m-M)$_{\rm LMC}$ = 18.477 $\pm$ 0.026 mag measured by \cite{Pietrzynski2019} leading to an agreement just within the margin of the errors. At this point it is hard to tell whether the difference is simply statistical, due to the depth of the LMC or caused by systematic effects not considered. Major potential sources for systematic effects are the model atmospheres used, their predicted spectra and energy distibutions and the calculation of synthetic photometry, which also involves the use of filter functions and photometric zero points. Our hope is that with more early type systems analyzed in exactly the same way it will be possible to identify potential systematic trends. This can then be used to calibrate the distance determinations with early type DEBs and then to use them for future applications at larger distances.

The surface brightnesses observed fit nicely on the extension to hotter stars of the surface brightness - color relationship obtained from interferometric measurements. Our goal is with the addition of more early type DEBs to obtain an accurate relationship at blue V-K$_s$ colors, which would then allow to determine surface brightnesses and distances without the model atmosphere dependent spectroscopic determination of effective temperatures.

In Paper I, we mentioned that for BLMC02 there was an inconsistency between the expected distance to the system, the measured reddening and the temperature from the $T_{eff}$-color relationship. 
In that study we finally adopted a temperature in accordance with the spectral type of the system. The results from the current analysis confirmed that our assumption regarding the distance and the reddening were correct and that indeed the temperature from $T_{eff}$-$(V-I)$ color calibration was wrong. This may indicate a problem with the calibration, which is very uncertain for hot stars, but also with the color that was used. As seen in Fig.~\ref{fig:colorfit}, the observed V-I color is bluer than the one from the model, which is consistent with higher temperature that we were obtaining in Paper I from this color.

The results of our test of stellar evolution are somewhat inconclusive. While we are able to produce evolutionary tracks, which reproduce the observed stellar parameters close to the margin of the errors, this requires a fine tuning of initial stellar rotational velocity and, most importantly, the internal transport of angular momentum. This procedure leads to an angular momentum transport, which is suspiciously low. Again, only the investigation of more early type DEB systems carried out in a similar way will show, whether this indicates a general problem or is just a peculiar result for one individual system. \\

\acknowledgments
MT gratefully acknowledges financial support from the Polish National Science Center grant PRELUDIUM 2016/21/N/ST9/03310.
This work was initiated during the Munich Institute for Astro and Particle Physics (MIAPP) 2018 program on the extragalactic distance scale. The stimulating atmosphere of this four week program and the support is gratefully acknowledged.
RPK has been supported by the Munich Excellence cluster ORIGINS funded by the Deutsche Forschungsgemeinschaft (DFG, German Research Foundation) under Germany's Excellence Strategy – EXC-2094 – 390783311.
We acknowledge support from the European Research Council (ERC) under the European Union's Horizon 2020 research and innovation program (grant agreement No 695099).
Support from the Polish National Science Centre grants MAESTRO UMO-2017/26/A/ST9/00446 and from the IdPII 2015 0002 64 grant of the Polish Ministry of Science and Higher Education is also acknowledged.
WG and GP acknowledge financial support from the BASAL Centro de Astrofisica y Tecnologias Afines (CATA, AFB-170002).  \\

\software{
	\texttt{MESA} \citep[v11701;][]{ Paxton2011, Paxton2013, Paxton2015, Paxton2018, Paxton2019}, \\}

\pagebreak
\bibliography{taormina_pII}

\begin{thebibliography}{}
\expandafter\ifx\csname natexlab\endcsname\relax\def\natexlab#1{#1}\fi
\providecommand{\url}[1]{\href{#1}{#1}}
\providecommand{\dodoi}[1]{doi:~\href{http://doi.org/#1}{\nolinkurl{#1}}}
\providecommand{\doeprint}[1]{\href{http://ascl.net/#1}{\nolinkurl{http://ascl.net/#1}}}
\providecommand{\doarXiv}[1]{\href{https://arxiv.org/abs/#1}{\nolinkurl{https://arxiv.org/abs/#1}}}

\bibitem[{{Bernstein} {et~al.}(2003){Bernstein}, {Shectman}, {Gunnels},
  {Mochnacki}, \& {Athey}}]{Bernstein2003}
{Bernstein}, R., {Shectman}, S.~A., {Gunnels}, S.~M., {Mochnacki}, S., \&
  {Athey}, A.~E. 2003, in \procspie, Vol. 4841, Instrument Design and
  Performance for Optical/Infrared Ground-based Telescopes, ed. M.~{Iye} \&
  A.~F.~M. {Moorwood}, 1694--1704

\bibitem[{{Bessell} \& {Murphy}(2012)}]{Bessell2012}
{Bessell}, M., \& {Murphy}, S. 2012, \pasp, 124, 140, \dodoi{10.1086/664083}

\bibitem[{{Bonanos} {et~al.}(2011){Bonanos}, {Castro}, {Macri}, \&
  {Kudritzki}}]{Bonanos2011}
{Bonanos}, A.~Z., {Castro}, N., {Macri}, L.~M., \& {Kudritzki}, R.-P. 2011,
  \apjl, 729, L9, \dodoi{10.1088/2041-8205/729/1/L9}

\bibitem[{{Bonanos} {et~al.}(2006){Bonanos}, {Stanek}, {Kudritzki}, {Macri},
  {Sasselov}, {Kaluzny}, {Stetson}, {Bersier}, {Bresolin}, \&
  {Matheson}}]{Bonanos2006}
{Bonanos}, A.~Z., {Stanek}, K.~Z., {Kudritzki}, R.~P., {et~al.} 2006, \apj,
  652, 313, \dodoi{10.1086/508140}

\bibitem[{{Brott} {et~al.}(2011){Brott}, {de Mink}, {Cantiello}, {Langer}, {de
  Koter}, {Evans}, {Hunter}, {Trundle}, \& {Vink}}]{Brott2011}
{Brott}, I., {de Mink}, S.~E., {Cantiello}, M., {et~al.} 2011, \aap, 530, A115,
  \dodoi{10.1051/0004-6361/201016113}

\bibitem[{{Cardelli} {et~al.}(1989){Cardelli}, {Clayton}, \&
  {Mathis}}]{Cardelli1989}
{Cardelli}, J.~A., {Clayton}, G.~C., \& {Mathis}, J.~S. 1989, \apj, 345, 245,
  \dodoi{10.1086/167900}

\bibitem[{{Challouf} {et~al.}(2014){Challouf}, {Nardetto}, {Mourard},
  {Graczyk}, {Aroui}, {Chesneau}, {Delaa}, {Pietrzy{\'n}ski}, {Gieren}, {Ligi},
  {Meilland }, {Perraut}, {Tallon-Bosc}, {McAlister}, {ten Brummelaar},
  {Sturmann}, {Sturmann}, {Turner}, {Farrington}, {Vargas}, \&
  {Scott}}]{Challouf2014}
{Challouf}, M., {Nardetto}, N., {Mourard}, D., {et~al.} 2014, \aap, 570, A104,
  \dodoi{10.1051/0004-6361/201423772}

\bibitem[{{Choi} {et~al.}(2016){Choi}, {Dotter}, {Conroy}, {Cantiello},
  {Paxton}, \& {Johnson}}]{Choi2016}
{Choi}, J., {Dotter}, A., {Conroy}, C., {et~al.} 2016, \apj, 823, 102,
  \dodoi{10.3847/0004-637X/823/2/102}

\bibitem[{{Cohen} {et~al.}(2003){Cohen}, {Wheaton}, \& {Megeath}}]{Cohen2003}
{Cohen}, M., {Wheaton}, W.~A., \& {Megeath}, S.~T. 2003, \aj, 126, 1090,
  \dodoi{10.1086/376474}

\bibitem[{{Cutri} {et~al.}(2012){Cutri}, {Skrutskie}, {van Dyk}, {Beichman},
  {Carpenter}, {Chester}, {Cambresy}, {Evans}, {Fowler}, {Gizis}, {Howard},
  {Huchra}, {Jarrett}, {Kopan}, {Kirkpatrick}, {Light}, {Marsh}, {McCallon},
  {Schneider}, {Stiening}, {Sykes}, {Weinberg}, {Wheaton}, {Wheelock}, \&
  {Zacharias}}]{Cutri2012}
{Cutri}, R.~M., {Skrutskie}, M.~F., {van Dyk}, S., {et~al.} 2012, VizieR Online
  Data Catalog, II/281

\bibitem[{{Dekker} {et~al.}(2000){Dekker}, {D'Odorico}, {Kaufer}, {Delabre}, \&
  {Kotzlowski}}]{Dekker2000}
{Dekker}, H., {D'Odorico}, S., {Kaufer}, A., {Delabre}, B., \& {Kotzlowski}, H.
  2000, in \procspie, Vol. 4008, Optical and IR Telescope Instrumentation and
  Detectors, ed. M.~{Iye} \& A.~F. {Moorwood}, 534--545

\bibitem[{{Di Benedetto}(2005)}]{DiBenedetto2005}
{Di Benedetto}, G.~P. 2005, \mnras, 357, 174,
  \dodoi{10.1111/j.1365-2966.2005.08632.x}

\bibitem[{{Ekstr{\"o}m} {et~al.}(2012){Ekstr{\"o}m}, {Georgy}, {Eggenberger},
  {Meynet}, {Mowlavi}, {Wyttenbach}, {Granada}, {Decressin}, {Hirschi},
  {Frischknecht}, {Charbonnel}, \& {Maeder}}]{Ekstrom2012}
{Ekstr{\"o}m}, S., {Georgy}, C., {Eggenberger}, P., {et~al.} 2012, \aap, 537,
  A146, \dodoi{10.1051/0004-6361/201117751}

\bibitem[{{Evans} {et~al.}(2018){Evans}, {Riello}, {De Angeli}, {Carrasco},
  {Montegriffo}, {Fabricius}, {Jordi}, {Palaversa}, {Diener}, {Busso},
  {Cacciari}, {van Leeuwen}, {Burgess}, {Davidson}, {Harrison}, {Hodgkin},
  {Pancino}, {Richards}, {Altavilla}, {Balaguer-N{\'u}{\~n}ez}, {Barstow},
  {Bellazzini}, {Brown}, {Castellani}, {Cocozza}, {De Luise}, {Delgado},
  {Ducourant}, {Galleti}, {Gilmore}, {Giuffrida}, {Holl}, {Kewley}, {Koposov},
  {Marinoni}, {Marrese}, {Osborne}, {Piersimoni}, {Portell}, {Pulone},
  {Ragaini}, {Sanna}, {Terrett}, {Walton}, {Wevers}, \&
  {Wyrzykowski}}]{Evans2018}
{Evans}, D.~W., {Riello}, M., {De Angeli}, F., {et~al.} 2018, \aap, 616, A4,
  \dodoi{10.1051/0004-6361/201832756}

\bibitem[{{Fitzpatrick}(1999)}]{Fitzpatrick1999}
{Fitzpatrick}, E.~L. 1999, \pasp, 111, 63, \dodoi{10.1086/316293}

\bibitem[{{Fitzpatrick} {et~al.}(2002){Fitzpatrick}, {Ribas}, {Guinan},
  {DeWarf}, {Maloney}, \& {Massa}}]{Fitzpatrick2002}
{Fitzpatrick}, E.~L., {Ribas}, I., {Guinan}, E.~F., {et~al.} 2002, \apj, 564,
  260, \dodoi{10.1086/324184}

\bibitem[{{Fitzpatrick} {et~al.}(2003){Fitzpatrick}, {Ribas}, {Guinan},
  {Maloney}, \& {Claret}}]{Fitzpatrick2003}
{Fitzpatrick}, E.~L., {Ribas}, I., {Guinan}, E.~F., {Maloney}, F.~P., \&
  {Claret}, A. 2003, \apj, 587, 685, \dodoi{10.1086/368309}

\bibitem[{{Freedman} {et~al.}(2019){Freedman}, {Madore}, {Hatt}, {Hoyt},
  {Jang}, {Beaton}, {Burns}, {Lee}, {Monson}, {Neeley}, {Phillips}, {Rich}, \&
  {Seibert}}]{Freedman2019}
{Freedman}, W.~L., {Madore}, B.~F., {Hatt}, D., {et~al.} 2019, arXiv e-prints,
  arXiv:1907.05922.
\newblock \doarXiv{1907.05922}

\bibitem[{{Gaia Collaboration} {et~al.}(2016){Gaia Collaboration}, {Prusti},
  {de Bruijne}, {Brown}, {Vallenari}, {Babusiaux}, {Bailer-Jones}, {Bastian},
  {Biermann}, {Evans}, {Eyer}, {Jansen}, {Jordi}, {Klioner}, {Lammers},
  {Lindegren}, {Luri}, {Mignard}, {Milligan}, {Panem}, {Poinsignon},
  {Pourbaix}, {Randich}, {Sarri}, {Sartoretti}, {Siddiqui}, {Soubiran},
  {Valette}, {van Leeuwen}, {Walton}, {Aerts}, {Arenou}, {Cropper}, {Drimmel},
  {H{\o}g}, {Katz}, {Lattanzi}, {O'Mullane}, {Grebel}, {Holland}, {Huc},
  {Passot}, {Bramante}, {Cacciari}, {Casta{\~n}eda}, {Chaoul}, {Cheek}, {De
  Angeli}, {Fabricius}, {Guerra}, {Hern{\'a}ndez}, {Jean-Antoine-Piccolo},
  {Masana}, {Messineo}, {Mowlavi}, {Nienartowicz}, {Ord{\'o}{\~n}ez-Blanco},
  {Panuzzo}, {Portell}, {Richards}, {Riello}, {Seabroke}, {Tanga},
  {Th{\'e}venin}, {Torra}, {Els}, {Gracia-Abril}, {Comoretto},
  {Garcia-Reinaldos}, {Lock}, {Mercier}, {Altmann}, {Andrae}, {Astraatmadja},
  {Bellas-Velidis}, {Benson}, {Berthier}, {Blomme}, {Busso}, {Carry},
  {Cellino}, {Clementini}, {Cowell}, {Creevey}, {Cuypers}, {Davidson}, {De
  Ridder}, {de Torres}, {Delchambre}, {Dell'Oro}, {Ducourant}, {Fr{\'e}mat},
  {Garc{\'\i}a-Torres}, {Gosset}, {Halbwachs}, {Hambly}, {Harrison}, {Hauser},
  {Hestroffer}, {Hodgkin}, {Huckle}, {Hutton}, {Jasniewicz}, {Jordan},
  {Kontizas}, {Korn}, {Lanzafame}, {Manteiga}, {Moitinho}, {Muinonen},
  {Osinde}, {Pancino}, {Pauwels}, {Petit}, {Recio-Blanco}, {Robin}, {Sarro},
  {Siopis}, {Smith}, {Smith}, {Sozzetti}, {Thuillot}, {van Reeven}, {Viala},
  {Abbas}, {Abreu Aramburu}, {Accart}, {Aguado}, {Allan}, {Allasia},
  {Altavilla}, {{\'A}lvarez}, {Alves}, {Anderson}, {Andrei}, {Anglada Varela},
  {Antiche}, {Antoja}, {Ant{\'o}n}, {Arcay}, {Atzei}, {Ayache}, {Bach},
  {Baker}, {Balaguer-N{\'u}{\~n}ez}, {Barache}, {Barata}, {Barbier}, {Barblan},
  {Baroni}, {Barrado y Navascu{\'e}s}, {Barros}, {Barstow}, {Becciani},
  {Bellazzini}, {Bellei}, {Bello Garc{\'\i}a}, {Belokurov}, {Bendjoya},
  {Berihuete}, {Bianchi}, {Bienaym{\'e}}, {Billebaud}, {Blagorodnova},
  {Blanco-Cuaresma}, {Boch}, {Bombrun}, {Borrachero}, {Bouquillon}, {Bourda},
  {Bouy}, {Bragaglia}, {Breddels}, {Brouillet}, {Br{\"u}semeister},
  {Bucciarelli}, {Budnik}, {Burgess}, {Burgon}, {Burlacu}, {Busonero}, {Buzzi},
  {Caffau}, {Cambras}, {Campbell}, {Cancelliere}, {Cantat-Gaudin}, {Carlucci},
  {Carrasco}, {Castellani}, {Charlot}, {Charnas}, {Charvet}, {Chassat},
  {Chiavassa}, {Clotet}, {Cocozza}, {Collins}, {Collins}, {Costigan}, {Crifo},
  {Cross}, {Crosta}, {Crowley}, {Dafonte}, {Damerdji}, {Dapergolas}, {David},
  {David}, {De Cat}, {de Felice}, {de Laverny}, {De Luise}, {De March}, {de
  Martino}, {de Souza}, {Debosscher}, {del Pozo}, {Delbo}, {Delgado},
  {Delgado}, {di Marco}, {Di Matteo}, {Diakite}, {Distefano}, {Dolding}, {Dos
  Anjos}, {Drazinos}, {Dur{\'a}n}, {Dzigan}, {Ecale}, {Edvardsson}, {Enke},
  {Erdmann}, {Escolar}, {Espina}, {Evans}, {Eynard Bontemps}, {Fabre},
  {Fabrizio}, {Faigler}, {Falc{\~a}o}, {Farr{\`a}s Casas}, {Faye}, {Federici},
  {Fedorets}, {Fern{\'a}ndez-Hern{\'a}ndez}, {Fernique}, {Fienga}, {Figueras},
  {Filippi}, {Findeisen}, {Fonti}, {Fouesneau}, {Fraile}, {Fraser}, {Fuchs},
  {Furnell}, {Gai}, {Galleti}, {Galluccio}, {Garabato}, {Garc{\'\i}a-Sedano},
  {Gar{\'e}}, {Garofalo}, {Garralda}, {Gavras}, {Gerssen}, {Geyer}, {Gilmore},
  {Girona}, {Giuffrida}, {Gomes}, {Gonz{\'a}lez-Marcos},
  {Gonz{\'a}lez-N{\'u}{\~n}ez}, {Gonz{\'a}lez-Vidal}, {Granvik}, {Guerrier},
  {Guillout}, {Guiraud}, {G{\'u}rpide}, {Guti{\'e}rrez-S{\'a}nchez}, {Guy},
  {Haigron}, {Hatzidimitriou}, {Haywood}, {Heiter}, {Helmi}, {Hobbs},
  {Hofmann}, {Holl}, {Holland }, {Hunt}, {Hypki}, {Icardi}, {Irwin}, {Jevardat
  de Fombelle}, {Jofr{\'e}}, {Jonker}, {Jorissen}, {Julbe}, {Karampelas},
  {Kochoska}, {Kohley}, {Kolenberg}, {Kontizas}, {Koposov}, {Kordopatis},
  {Koubsky}, {Kowalczyk}, {Krone-Martins}, {Kudryashova}, {Kull}, {Bachchan},
  {Lacoste-Seris}, {Lanza}, {Lavigne}, {Le Poncin-Lafitte}, {Lebreton},
  {Lebzelter}, {Leccia}, {Leclerc}, {Lecoeur-Taibi}, {Lemaitre}, {Lenhardt},
  {Leroux}, {Liao}, {Licata}, {Lindstr{\o}m}, {Lister}, {Livanou}, {Lobel},
  {L{\"o}ffler}, {L{\'o}pez}, {Lopez-Lozano}, {Lorenz}, {Loureiro},
  {MacDonald}, {Magalh{\~a}es Fernandes}, {Managau}, {Mann}, {Mantelet},
  {Marchal}, {Marchant}, {Marconi}, {Marie}, {Marinoni}, {Marrese},
  {Marschalk{\'o}}, {Marshall}, {Mart{\'\i}n-Fleitas}, {Martino}, {Mary},
  {Matijevi{\v{c}}}, {Mazeh}, {McMillan}, {Messina}, {Mestre}, {Michalik},
  {Millar}, {Miranda}, {Molina}, {Molinaro}, {Molinaro}, {Moln{\'a}r},
  {Moniez}, {Montegriffo}, {Monteiro}, {Mor}, {Mora}, {Morbidelli}, {Morel},
  {Morgenthaler}, {Morley}, {Morris}, {Mulone}, {Muraveva}, {Musella},
  {Narbonne}, {Nelemans}, {Nicastro}, {Noval}, {Ord{\'e}novic},
  {Ordieres-Mer{\'e}}, {Osborne}, {Pagani}, {Pagano}, {Pailler}, {Palacin},
  {Palaversa}, {Parsons}, {Paulsen}, {Pecoraro}, {Pedrosa}, {Pentik{\"a}inen},
  {Pereira}, {Pichon}, {Piersimoni}, {Pineau}, {Plachy}, {Plum}, {Poujoulet},
  {Pr{\v{s}}a}, {Pulone}, {Ragaini}, {Rago}, {Rambaux}, {Ramos-Lerate},
  {Ranalli}, {Rauw}, {Read}, {Regibo}, {Renk}, {Reyl{\'e}}, {Ribeiro},
  {Rimoldini}, {Ripepi}, {Riva}, {Rixon}, {Roelens}, {Romero-G{\'o}mez},
  {Rowell}, {Royer}, {Rudolph}, {Ruiz-Dern}, {Sadowski}, {Sagrist{\`a}
  Sell{\'e}s}, {Sahlmann}, {Salgado}, {Salguero}, {Sarasso}, {Savietto},
  {Schnorhk}, {Schultheis}, {Sciacca}, {Segol}, {Segovia}, {Segransan},
  {Serpell}, {Shih}, {Smareglia}, {Smart}, {Smith}, {Solano}, {Solitro},
  {Sordo}, {Soria Nieto}, {Souchay}, {Spagna}, {Spoto}, {Stampa}, {Steele},
  {Steidelm{\"u}ller}, {Stephenson}, {Stoev}, {Suess}, {S{\"u}veges}, {Surdej},
  {Szabados}, {Szegedi-Elek}, {Tapiador}, {Taris}, {Tauran}, {Taylor},
  {Teixeira}, {Terrett}, {Tingley}, {Trager}, {Turon}, {Ulla}, {Utrilla},
  {Valentini}, {van Elteren}, {Van Hemelryck}, {van Leeuwen}, {Varadi},
  {Vecchiato}, {Veljanoski}, {Via}, {Vicente}, {Vogt}, {Voss}, {Votruba},
  {Voutsinas}, {Walmsley}, {Weiler}, {Weingrill}, {Werner}, {Wevers},
  {Whitehead}, {Wyrzykowski}, {Yoldas}, {{\v{Z}}erjal}, {Zucker}, {Zurbach},
  {Zwitter}, {Alecu}, {Allen}, {Allende Prieto}, {Amorim},
  {Anglada-Escud{\'e}}, {Arsenijevic}, {Azaz}, {Balm}, {Beck}, {Bernstein},
  {Bigot}, {Bijaoui}, {Blasco}, {Bonfigli}, {Bono}, {Boudreault}, {Bressan},
  {Brown}, {Brunet}, {Bunclark}, {Buonanno}, {Butkevich}, {Carret}, {Carrion},
  {Chemin}, {Ch{\'e}reau}, {Corcione}, {Darmigny}, {de Boer}, {de Teodoro}, {de
  Zeeuw}, {Delle Luche}, {Domingues}, {Dubath}, {Fodor}, {Fr{\'e}zouls},
  {Fries}, {Fustes}, {Fyfe}, {Gallardo}, {Gallegos}, {Gardiol}, {Gebran},
  {Gomboc}, {G{\'o}mez}, {Grux}, {Gueguen}, {Heyrovsky}, {Hoar}, {Iannicola},
  {Isasi Parache}, {Janotto}, {Joliet}, {Jonckheere}, {Keil}, {Kim},
  {Klagyivik}, {Klar}, {Knude}, {Kochukhov}, {Kolka}, {Kos}, {Kutka}, {Lainey},
  {LeBouquin}, {Liu}, {Loreggia}, {Makarov}, {Marseille}, {Martayan},
  {Martinez-Rubi}, {Massart}, {Meynadier}, {Mignot}, {Munari}, {Nguyen},
  {Nordlander}, {Ocvirk}, {O'Flaherty}, {Olias Sanz}, {Ortiz}, {Osorio},
  {Oszkiewicz}, {Ouzounis}, {Palmer}, {Park}, {Pasquato}, {Peltzer}, {Peralta},
  {P{\'e}turaud}, {Pieniluoma}, {Pigozzi}, {Poels}, {Prat}, {Prod'homme},
  {Raison}, {Rebordao}, {Risquez}, {Rocca-Volmerange}, {Rosen}, {Ruiz-Fuertes},
  {Russo}, {Sembay}, {Serraller Vizcaino}, {Short}, {Siebert}, {Silva},
  {Sinachopoulos}, {Slezak}, {Soffel}, {Sosnowska}, {Strai{\v{z}}ys}, {ter
  Linden}, {Terrell}, {Theil}, {Tiede}, {Troisi}, {Tsalmantza}, {Tur},
  {Vaccari}, {Vachier}, {Valles}, {Van Hamme}, {Veltz}, {Virtanen}, {Wallut},
  {Wichmann}, {Wilkinson}, {Ziaeepour}, \& {Zschocke}}]{Gaia2016}
{Gaia Collaboration}, {Prusti}, T., {de Bruijne}, J.~H.~J., {et~al.} 2016,
  \aap, 595, A1, \dodoi{10.1051/0004-6361/201629272}

\bibitem[{{Gaia Collaboration} {et~al.}(2018){Gaia Collaboration}, {Brown},
  {Vallenari}, {Prusti}, {de Bruijne}, {Babusiaux}, {Bailer-Jones}, {Biermann},
  {Evans}, {Eyer}, {Jansen}, {Jordi}, {Klioner}, {Lammers}, {Lindegren},
  {Luri}, {Mignard}, {Panem}, {Pourbaix}, {Randich}, {Sartoretti}, {Siddiqui},
  {Soubiran}, {van Leeuwen}, {Walton}, {Arenou}, {Bastian}, {Cropper},
  {Drimmel}, {Katz}, {Lattanzi}, {Bakker}, {Cacciari}, {Casta{\~n}eda},
  {Chaoul}, {Cheek}, {De Angeli}, {Fabricius}, {Guerra}, {Holl}, {Masana},
  {Messineo}, {Mowlavi}, {Nienartowicz}, {Panuzzo}, {Portell}, {Riello},
  {Seabroke}, {Tanga}, {Th{\'e}venin}, {Gracia-Abril}, {Comoretto},
  {Garcia-Reinaldos}, {Teyssier}, {Altmann}, {Andrae}, {Audard},
  {Bellas-Velidis}, {Benson}, {Berthier}, {Blomme}, {Burgess}, {Busso},
  {Carry}, {Cellino}, {Clementini}, {Clotet}, {Creevey}, {Davidson}, {De
  Ridder}, {Delchambre}, {Dell'Oro}, {Ducourant},
  {Fern{\'a}ndez-Hern{\'a}ndez}, {Fouesneau}, {Fr{\'e}mat}, {Galluccio},
  {Garc{\'\i}a-Torres}, {Gonz{\'a}lez-N{\'u}{\~n}ez}, {Gonz{\'a}lez-Vidal},
  {Gosset}, {Guy}, {Halbwachs}, {Hambly}, {Harrison}, {Hern{\'a}ndez},
  {Hestroffer}, {Hodgkin}, {Hutton}, {Jasniewicz}, {Jean-Antoine-Piccolo},
  {Jordan}, {Korn}, {Krone-Martins}, {Lanzafame}, {Lebzelter}, {L{\"o}ffler},
  {Manteiga}, {Marrese}, {Mart{\'\i}n-Fleitas}, {Moitinho}, {Mora}, {Muinonen},
  {Osinde}, {Pancino}, {Pauwels}, {Petit}, {Recio-Blanco}, {Richards},
  {Rimoldini}, {Robin}, {Sarro}, {Siopis}, {Smith}, {Sozzetti}, {S{\"u}veges},
  {Torra}, {van Reeven}, {Abbas}, {Abreu Aramburu}, {Accart}, {Aerts},
  {Altavilla}, {{\'A}lvarez}, {Alvarez}, {Alves}, {Anderson}, {Andrei},
  {Anglada Varela}, {Antiche}, {Antoja}, {Arcay}, {Astraatmadja}, {Bach},
  {Baker}, {Balaguer-N{\'u}{\~n}ez}, {Balm}, {Barache}, {Barata}, {Barbato},
  {Barblan}, {Barklem}, {Barrado}, {Barros}, {Barstow}, {Bartholom{\'e}
  Mu{\~n}oz}, {Bassilana}, {Becciani}, {Bellazzini}, {Berihuete}, {Bertone},
  {Bianchi}, {Bienaym{\'e}}, {Blanco-Cuaresma}, {Boch}, {Boeche}, {Bombrun},
  {Borrachero}, {Bossini}, {Bouquillon}, {Bourda}, {Bragaglia}, {Bramante},
  {Breddels}, {Bressan}, {Brouillet}, {Br{\"u}semeister}, {Brugaletta},
  {Bucciarelli}, {Burlacu}, {Busonero}, {Butkevich}, {Buzzi}, {Caffau},
  {Cancelliere}, {Cannizzaro}, {Cantat-Gaudin}, {Carballo}, {Carlucci},
  {Carrasco}, {Casamiquela}, {Castellani}, {Castro-Ginard}, {Charlot},
  {Chemin}, {Chiavassa}, {Cocozza}, {Costigan}, {Cowell}, {Crifo}, {Crosta},
  {Crowley}, {Cuypers}, {Dafonte}, {Damerdji}, {Dapergolas}, {David}, {David},
  {de Laverny}, {De Luise}, {De March}, {de Martino}, {de Souza}, {de Torres},
  {Debosscher}, {del Pozo}, {Delbo}, {Delgado}, {Delgado}, {Di Matteo},
  {Diakite}, {Diener}, {Distefano}, {Dolding}, {Drazinos}, {Dur{\'a}n},
  {Edvardsson}, {Enke}, {Eriksson}, {Esquej}, {Eynard Bontemps}, {Fabre},
  {Fabrizio}, {Faigler}, {Falc{\~a}o}, {Farr{\`a}s Casas}, {Federici},
  {Fedorets}, {Fernique}, {Figueras}, {Filippi}, {Findeisen}, {Fonti},
  {Fraile}, {Fraser}, {Fr{\'e}zouls}, {Gai}, {Galleti}, {Garabato},
  {Garc{\'\i}a-Sedano}, {Garofalo}, {Garralda}, {Gavel}, {Gavras}, {Gerssen},
  {Geyer}, {Giacobbe}, {Gilmore}, {Girona}, {Giuffrida}, {Glass}, {Gomes},
  {Granvik}, {Gueguen}, {Guerrier}, {Guiraud}, {Guti{\'e}rrez-S{\'a}nchez},
  {Haigron}, {Hatzidimitriou}, {Hauser}, {Haywood}, {Heiter}, {Helmi}, {Heu},
  {Hilger}, {Hobbs}, {Hofmann}, {Holland}, {Huckle}, {Hypki}, {Icardi},
  {Jan{\ss}en}, {Jevardat de Fombelle}, {Jonker}, {Juh{\'a}sz}, {Julbe},
  {Karampelas}, {Kewley}, {Klar}, {Kochoska}, {Kohley}, {Kolenberg},
  {Kontizas}, {Kontizas}, {Koposov}, {Kordopatis}, {Kostrzewa-Rutkowska},
  {Koubsky}, {Lambert}, {Lanza}, {Lasne}, {Lavigne}, {Le Fustec}, {Le
  Poncin-Lafitte}, {Lebreton}, {Leccia}, {Leclerc}, {Lecoeur-Taibi},
  {Lenhardt}, {Leroux}, {Liao}, {Licata}, {Lindstr{\o}m}, {Lister}, {Livanou},
  {Lobel}, {L{\'o}pez}, {Managau}, {Mann}, {Mantelet}, {Marchal}, {Marchant},
  {Marconi}, {Marinoni}, {Marschalk{\'o}}, {Marshall}, {Martino}, {Marton},
  {Mary}, {Massari}, {Matijevi{\v{c}}}, {Mazeh}, {McMillan}, {Messina},
  {Michalik}, {Millar}, {Molina}, {Molinaro}, {Moln{\'a}r}, {Montegriffo},
  {Mor}, {Morbidelli}, {Morel}, {Morris}, {Mulone}, {Muraveva}, {Musella},
  {Nelemans}, {Nicastro}, {Noval}, {O'Mullane}, {Ord{\'e}novic},
  {Ord{\'o}{\~n}ez-Blanco}, {Osborne}, {Pagani}, {Pagano}, {Pailler},
  {Palacin}, {Palaversa}, {Panahi}, {Pawlak}, {Piersimoni}, {Pineau}, {Plachy},
  {Plum}, {Poggio}, {Poujoulet}, {Pr{\v{s}}a}, {Pulone}, {Racero}, {Ragaini},
  {Rambaux}, {Ramos-Lerate}, {Regibo}, {Reyl{\'e}}, {Riclet}, {Ripepi}, {Riva},
  {Rivard}, {Rixon}, {Roegiers}, {Roelens}, {Romero-G{\'o}mez}, {Rowell},
  {Royer}, {Ruiz-Dern}, {Sadowski}, {Sagrist{\`a} Sell{\'e}s}, {Sahlmann},
  {Salgado}, {Salguero}, {Sanna}, {Santana-Ros}, {Sarasso}, {Savietto},
  {Schultheis}, {Sciacca}, {Segol}, {Segovia}, {S{\'e}gransan}, {Shih},
  {Siltala}, {Silva}, {Smart}, {Smith}, {Solano}, {Solitro}, {Sordo}, {Soria
  Nieto}, {Souchay}, {Spagna}, {Spoto}, {Stampa}, {Steele},
  {Steidelm{\"u}ller}, {Stephenson}, {Stoev}, {Suess}, {Surdej}, {Szabados},
  {Szegedi-Elek}, {Tapiador}, {Taris}, {Tauran}, {Taylor}, {Teixeira},
  {Terrett}, {Teyssand ier}, {Thuillot}, {Titarenko}, {Torra Clotet}, {Turon},
  {Ulla}, {Utrilla}, {Uzzi}, {Vaillant}, {Valentini}, {Valette}, {van Elteren},
  {Van Hemelryck}, {van Leeuwen}, {Vaschetto}, {Vecchiato}, {Veljanoski},
  {Viala}, {Vicente}, {Vogt}, {von Essen}, {Voss}, {Votruba}, {Voutsinas},
  {Walmsley}, {Weiler}, {Wertz}, {Wevers}, {Wyrzykowski}, {Yoldas},
  {{\v{Z}}erjal}, {Ziaeepour}, {Zorec}, {Zschocke}, {Zucker}, {Zurbach}, \&
  {Zwitter}}]{Gaia2018}
{Gaia Collaboration}, {Brown}, A.~G.~A., {Vallenari}, A., {et~al.} 2018, \aap,
  616, A1, \dodoi{10.1051/0004-6361/201833051}

\bibitem[{{Georgy} {et~al.}(2013){Georgy}, {Ekstr{\"o}m}, {Eggenberger},
  {Meynet}, {Haemmerl{\'e}}, {Maeder}, {Granada}, {Groh}, {Hirschi}, {Mowlavi},
  {Yusof}, {Charbonnel}, {Decressin}, \& {Barblan}}]{Georgy2013}
{Georgy}, C., {Ekstr{\"o}m}, S., {Eggenberger}, P., {et~al.} 2013, \aap, 558,
  A103, \dodoi{10.1051/0004-6361/201322178}

\bibitem[{{Groh} {et~al.}(2019){Groh}, {Ekstr{\"o}m}, {Georgy}, {Meynet},
  {Choplin}, {Eggenberger}, {Hirschi}, {Maeder}, {Murphy}, {Boian}, \&
  {Farrell}}]{Groh2019}
{Groh}, J.~H., {Ekstr{\"o}m}, S., {Georgy}, C., {et~al.} 2019, \aap, 627, A24,
  \dodoi{10.1051/0004-6361/201833720}

\bibitem[{{Guinan} {et~al.}(1998){Guinan}, {Fitzpatrick}, {DeWarf}, {Maloney},
  {Maurone}, {Ribas}, {Pritchard}, {Bradstreet}, \& {Gim{\'e}nez}}]{Guinan1998}
{Guinan}, E.~F., {Fitzpatrick}, E.~L., {DeWarf}, L.~E., {et~al.} 1998, \apjl,
  509, L21, \dodoi{10.1086/311760}

\bibitem[{{Heger} {et~al.}(2000){Heger}, {Langer}, \& {Woosley}}]{Heger2000}
{Heger}, A., {Langer}, N., \& {Woosley}, S.~E. 2000, \apj, 528, 368,
  \dodoi{10.1086/308158}

\bibitem[{{Herrero} {et~al.}(1992){Herrero}, {Kudritzki}, {Vilchez}, {Kunze},
  {Butler}, \& {Haser}}]{Herrero1992}
{Herrero}, A., {Kudritzki}, R.~P., {Vilchez}, J.~M., {et~al.} 1992, \aap, 261,
  209

\bibitem[{{Holgado} {et~al.}(2018){Holgado}, {Sim{\'o}n-D{\'\i}az},
  {Barb{\'a}}, {Puls}, {Herrero}, {Castro}, {Garcia}, {Ma{\'\i}z
  Apell{\'a}niz}, {Negueruela}, \& {Sab{\'\i}n-Sanjuli{\'a}n}}]{Holgado2018}
{Holgado}, G., {Sim{\'o}n-D{\'\i}az}, S., {Barb{\'a}}, R.~H., {et~al.} 2018,
  \aap, 613, A65, \dodoi{10.1051/0004-6361/201731543}

\bibitem[{{Hunter} {et~al.}(2008){Hunter}, {Brott}, {Lennon}, {Langer},
  {Dufton}, {Trundle}, {Smartt}, {de Koter}, {Evans}, \& {Ryans}}]{Hunter2008}
{Hunter}, I., {Brott}, I., {Lennon}, D.~J., {et~al.} 2008, \apjl, 676, L29,
  \dodoi{10.1086/587436}

\bibitem[{{Josselin} \& {Plez}(2007)}]{Josselin2007}
{Josselin}, E., \& {Plez}, B. 2007, \aap, 469, 671,
  \dodoi{10.1051/0004-6361:20066353}

\bibitem[{{Komatsu} {et~al.}(2011){Komatsu}, {Smith}, {Dunkley}, {Bennett},
  {Gold}, {Hinshaw}, {Jarosik}, {Larson}, {Nolta}, \& {Page}}]{Komatsu2011}
{Komatsu}, E., {Smith}, K.~M., {Dunkley}, J., {et~al.} 2011, \apjs, 192, 18,
  \dodoi{10.1088/0067-0049/192/2/18}

\bibitem[{{Kudritzki} \& {Puls}(2000)}]{Kudritzki2000}
{Kudritzki}, R.-P., \& {Puls}, J. 2000, \araa, 38, 613,
  \dodoi{10.1146/annurev.astro.38.1.613}

\bibitem[{{Kudritzki} {et~al.}(2006){Kudritzki}, {Urbaneja}, \&
  {Puls}}]{Kudritzki2006}
{Kudritzki}, R.~P., {Urbaneja}, M.~A., \& {Puls}, J. 2006, in IAU Symposium,
  Vol. 234, Planetary Nebulae in our Galaxy and Beyond, ed. M.~J. {Barlow} \&
  R.~H. {M{\'e}ndez}, 119--126

\bibitem[{{Ma{\'\i}z Apell{\'a}niz} {et~al.}(2017){Ma{\'\i}z Apell{\'a}niz},
  {Trigueros P{\'a}ez}, {Bostroem}, {Barb{\'a}}, \& {Evans}}]{Maiz2017}
{Ma{\'\i}z Apell{\'a}niz}, J., {Trigueros P{\'a}ez}, E., {Bostroem}, A.~K.,
  {Barb{\'a}}, R.~H., \& {Evans}, C.~J. 2017, in Highlights on Spanish
  Astrophysics IX, 510--510

\bibitem[{{Ma{\'\i}z Apell{\'a}niz} {et~al.}(2014){Ma{\'\i}z Apell{\'a}niz},
  {Evans}, {Barb{\'a}}, {Gr{\"a}fener}, {Bestenlehner}, {Crowther},
  {Garc{\'\i}a}, {Herrero}, {Sana}, \& {Sim{\'o}n-D{\'\i}az}}]{Maiz2014}
{Ma{\'\i}z Apell{\'a}niz}, J., {Evans}, C.~J., {Barb{\'a}}, R.~H., {et~al.}
  2014, \aap, 564, A63, \dodoi{10.1051/0004-6361/201423439}

\bibitem[{{Markova} {et~al.}(2018){Markova}, {Puls}, \& {Langer}}]{Markova2018}
{Markova}, N., {Puls}, J., \& {Langer}, N. 2018, \aap, 613, A12,
  \dodoi{10.1051/0004-6361/201731361}

\bibitem[{{Martins} {et~al.}(2015){Martins}, {Herv{\'e}}, {Bouret},
  {Marcolino}, {Wade}, {Neiner}, {Alecian}, {Grunhut}, \&
  {Petit}}]{Martins2015}
{Martins}, F., {Herv{\'e}}, A., {Bouret}, J.~C., {et~al.} 2015, \aap, 575, A34,
  \dodoi{10.1051/0004-6361/201425173}

\bibitem[{{Nelson} {et~al.}(2000){Nelson}, {Cook}, {Popowski}, \&
  {Alves}}]{Nelson2000}
{Nelson}, C.~A., {Cook}, K.~H., {Popowski}, P., \& {Alves}, D.~R. 2000, \aj,
  119, 1205, \dodoi{10.1086/301268}

\bibitem[{{O'Donnell}(1994)}]{Odonnell1994}
{O'Donnell}, J.~E. 1994, \apj, 422, 158, \dodoi{10.1086/173713}

\bibitem[{{Paxton} {et~al.}(2011){Paxton}, {Bildsten}, {Dotter}, {Herwig},
  {Lesaffre}, \& {Timmes}}]{Paxton2011}
{Paxton}, B., {Bildsten}, L., {Dotter}, A., {et~al.} 2011, \apjs, 192, 3,
  \dodoi{10.1088/0067-0049/192/1/3}

\bibitem[{{Paxton} {et~al.}(2013){Paxton}, {Cantiello}, {Arras}, {Bildsten},
  {Brown}, {Dotter}, {Mankovich}, {Montgomery}, {Stello}, {Timmes}, \&
  {Townsend}}]{Paxton2013}
{Paxton}, B., {Cantiello}, M., {Arras}, P., {et~al.} 2013, \apjs, 208, 4,
  \dodoi{10.1088/0067-0049/208/1/4}

\bibitem[{{Paxton} {et~al.}(2015){Paxton}, {Marchant}, {Schwab}, {Bauer},
  {Bildsten}, {Cantiello}, {Dessart}, {Farmer}, {Hu}, {Langer}, {Townsend},
  {Townsley}, \& {Timmes}}]{Paxton2015}
{Paxton}, B., {Marchant}, P., {Schwab}, J., {et~al.} 2015, \apjs, 220, 15,
  \dodoi{10.1088/0067-0049/220/1/15}

\bibitem[{{Paxton} {et~al.}(2018){Paxton}, {Schwab}, {Bauer}, {Bildsten},
  {Blinnikov}, {Duffell}, {Farmer}, {Goldberg}, {Marchant}, {Sorokina},
  {Thoul}, {Townsend}, \& {Timmes}}]{Paxton2018}
{Paxton}, B., {Schwab}, J., {Bauer}, E.~B., {et~al.} 2018, \apjs, 234, 34,
  \dodoi{10.3847/1538-4365/aaa5a8}

\bibitem[{{Paxton} {et~al.}(2019){Paxton}, {Smolec}, {Schwab}, {Gautschy},
  {Bildsten}, {Cantiello}, {Dotter}, {Farmer}, {Goldberg}, {Jermyn}, {Kanbur},
  {Marchant}, {Thoul}, {Townsend}, {Wolf}, {Zhang}, \& {Timmes}}]{Paxton2019}
{Paxton}, B., {Smolec}, R., {Schwab}, J., {et~al.} 2019, \apjs, 243, 10,
  \dodoi{10.3847/1538-4365/ab2241}

\bibitem[{{Pietrzy{\'n}ski} {et~al.}(2019){Pietrzy{\'n}ski}, {Graczyk},
  {Gallenne}, {Gieren}, {Thompson}, {Pilecki}, {Karczmarek}, {G{\'o}rski},
  {Suchomska}, \& {Taormina}}]{Pietrzynski2019}
{Pietrzy{\'n}ski}, G., {Graczyk}, D., {Gallenne}, A., {et~al.} 2019, \nat, 567,
  200, \dodoi{10.1038/s41586-019-0999-4}

\bibitem[{{Puls} {et~al.}(2006){Puls}, {Markova}, {Scuderi}, {Stanghellini},
  {Taranova}, {Burnley}, \& {Howarth}}]{Puls2006}
{Puls}, J., {Markova}, N., {Scuderi}, S., {et~al.} 2006, \aap, 454, 625,
  \dodoi{10.1051/0004-6361:20065073}

\bibitem[{{Puls} {et~al.}(2009){Puls}, {Sundqvist}, {Najarro}, \&
  {Hanson}}]{Puls2009}
{Puls}, J., {Sundqvist}, J.~O., {Najarro}, F., \& {Hanson}, M.~M. 2009, in
  American Institute of Physics Conference Series, Vol. 1171, American
  Institute of Physics Conference Series, ed. I.~{Hubeny}, J.~M. {Stone},
  K.~{MacGregor}, \& K.~{Werner}, 123--135

\bibitem[{{Puls} {et~al.}(2005){Puls}, {Urbaneja}, {Venero}, {Repolust},
  {Springmann}, {Jokuthy}, \& {Mokiem}}]{Puls2005}
{Puls}, J., {Urbaneja}, M.~A., {Venero}, R., {et~al.} 2005, \aap, 435, 669,
  \dodoi{10.1051/0004-6361:20042365}

\bibitem[{{Puls} {et~al.}(2008){Puls}, {Vink}, \& {Najarro}}]{Puls2008}
{Puls}, J., {Vink}, J.~S., \& {Najarro}, F. 2008, \aapr, 16, 209,
  \dodoi{10.1007/s00159-008-0015-8}

\bibitem[{{Puls} {et~al.}(1996){Puls}, {Kudritzki}, {Herrero}, {Pauldrach},
  {Haser}, {Lennon}, {Gabler}, {Voels}, {Vilchez}, \& {Wachter}}]{Puls1996}
{Puls}, J., {Kudritzki}, R.~P., {Herrero}, A., {et~al.} 1996, \aap, 305, 171

\bibitem[{{Ribas} {et~al.}(2002){Ribas}, {Fitzpatrick}, {Maloney}, {Guinan}, \&
  {Udalski}}]{Ribas2002}
{Ribas}, I., {Fitzpatrick}, E.~L., {Maloney}, F.~P., {Guinan}, E.~F., \&
  {Udalski}, A. 2002, \apj, 574, 771, \dodoi{10.1086/341003}

\bibitem[{{Ribas} {et~al.}(2005){Ribas}, {Jordi}, {Vilardell}, {Fitzpatrick},
  {Hilditch}, \& {Guinan}}]{Ribas2005}
{Ribas}, I., {Jordi}, C., {Vilardell}, F., {et~al.} 2005, \apjl, 635, L37,
  \dodoi{10.1086/499161}

\bibitem[{{Ribas} {et~al.}(2000){Ribas}, {Guinan}, {Fitzpatrick}, {DeWarf},
  {Maloney}, {Maurone}, {Bradstreet}, {Gim{\'e}nez}, \&
  {Pritchard}}]{Ribas2000}
{Ribas}, I., {Guinan}, E.~F., {Fitzpatrick}, E.~L., {et~al.} 2000, \apj, 528,
  692, \dodoi{10.1086/308210}

\bibitem[{{Riess} {et~al.}(2019){Riess}, {Casertano}, {Yuan}, {Macri}, \&
  {Scolnic}}]{Riess2019}
{Riess}, A.~G., {Casertano}, S., {Yuan}, W., {Macri}, L.~M., \& {Scolnic}, D.
  2019, \apj, 876, 85, \dodoi{10.3847/1538-4357/ab1422}

\bibitem[{{Rivero Gonz{\'a}lez} {et~al.}(2012{\natexlab{a}}){Rivero
  Gonz{\'a}lez}, {Puls}, {Massey}, \& {Najarro}}]{Rivero2012b}
{Rivero Gonz{\'a}lez}, J.~G., {Puls}, J., {Massey}, P., \& {Najarro}, F.
  2012{\natexlab{a}}, \aap, 543, A95, \dodoi{10.1051/0004-6361/201218955}

\bibitem[{{Rivero Gonz{\'a}lez} {et~al.}(2012{\natexlab{b}}){Rivero
  Gonz{\'a}lez}, {Puls}, {Najarro}, \& {Brott}}]{Rivero2012a}
{Rivero Gonz{\'a}lez}, J.~G., {Puls}, J., {Najarro}, F., \& {Brott}, I.
  2012{\natexlab{b}}, \aap, 537, A79, \dodoi{10.1051/0004-6361/201117790}

\bibitem[{{Spruit}(2002)}]{Spruit2002}
{Spruit}, H.~C. 2002, \aap, 381, 923, \dodoi{10.1051/0004-6361:20011465}

\bibitem[{{Taormina} {et~al.}(2019){Taormina}, {Pietrzy{\'n}ski}, {Pilecki},
  {Kudritzki}, {Thompson}, {Graczyk}, {Gieren}, {Nardetto}, {G{\'o}rski},
  {Suchomska}, {Zgirski}, {Wielg{\'o}rski}, {Karczmarek}, \&
  {Narloch}}]{Taormina2019}
{Taormina}, M., {Pietrzy{\'n}ski}, G., {Pilecki}, B., {et~al.} 2019, \apj, 886,
  111, \dodoi{10.3847/1538-4357/ab4b57}

\bibitem[{{Udalski} {et~al.}(1998){Udalski}, {Pietrzy{\'n}ski}, {Wo{\'z}niak},
  {Szyma{\'n}ski}, {Kubiak}, \& {{\.Z}ebru{\'n} }}]{Udalski1998}
{Udalski}, A., {Pietrzy{\'n}ski}, G., {Wo{\'z}niak}, P., {et~al.} 1998, \apjl,
  509, L25, \dodoi{10.1086/311763}

\bibitem[{{Urbaneja} {et~al.}(2017){Urbaneja}, {Kudritzki}, {Gieren},
  {Pietrzy{\'n}ski}, {Bresolin}, \& {Przybilla}}]{Urbaneja2017}
{Urbaneja}, M.~A., {Kudritzki}, R.-P., {Gieren}, W., {et~al.} 2017, \aj, 154,
  102, \dodoi{10.3847/1538-3881/aa79a8}

\bibitem[{{van der Marel} \& {Kallivayalil}(2014)}]{vanderMarel2014}
{van der Marel}, R.~P., \& {Kallivayalil}, N. 2014, \apj, 781, 121,
  \dodoi{10.1088/0004-637X/781/2/121}

\bibitem[{{Vilardell} {et~al.}(2010){Vilardell}, {Ribas}, {Jordi},
  {Fitzpatrick}, \& {Guinan}}]{Vilardell2010}
{Vilardell}, F., {Ribas}, I., {Jordi}, C., {Fitzpatrick}, E.~L., \& {Guinan},
  E.~F. 2010, \aap, 509, A70, \dodoi{10.1051/0004-6361/200913299}

\bibitem[{{Vink} {et~al.}(1999){Vink}, {de Koter}, \& {Lamers}}]{Vink1999}
{Vink}, J.~S., {de Koter}, A., \& {Lamers}, H.~J.~G.~L.~M. 1999, \aap, 350,
  181.
\newblock \doarXiv{astro-ph/9908196}

\bibitem[{{Vink} {et~al.}(2000){Vink}, {de Koter}, \& {Lamers}}]{Vink2000}
---. 2000, \aap, 362, 295.
\newblock \doarXiv{astro-ph/0008183}

\bibitem[{{Vink} {et~al.}(2001){Vink}, {de Koter}, \& {Lamers}}]{Vink2001}
---. 2001, \aap, 369, 574, \dodoi{10.1051/0004-6361:20010127}

\bibitem[{{Weinberg} {et~al.}(2013){Weinberg}, {Mortonson}, {Eisenstein},
  {Hirata}, {Riess}, \& {Rozo}}]{Weinberg2013}
{Weinberg}, D.~H., {Mortonson}, M.~J., {Eisenstein}, D.~J., {et~al.} 2013,
  \physrep, 530, 87, \dodoi{10.1016/j.physrep.2013.05.001}

\end{thebibliography}

\end{document}